\documentclass[pdflatex,sn-mathphys-num]{sn-jnl}

\usepackage[T1]{fontenc}
\usepackage{graphicx}
\usepackage{multirow}
\usepackage{amsmath,amssymb,amsfonts}
\usepackage{amsthm}
\usepackage{mathrsfs}
\usepackage[title]{appendix}
\usepackage{xcolor}
\usepackage{pdfpages}
\usepackage{textcomp}
\usepackage{manyfoot}
\usepackage{booktabs}
\usepackage{algorithm}
\usepackage{algorithmicx}
\usepackage{algpseudocode}
\usepackage{listings}
\usepackage{float}
\usepackage{subcaption}

\renewcommand{\topfraction}{0.95}
\renewcommand{\bottomfraction}{0.0}
\renewcommand{\textfraction}{0.05}
\renewcommand{\floatpagefraction}{0.7}

\theoremstyle{thmstyleone}

\theoremstyle{thmstyletwo}

\theoremstyle{thmstylethree}

\begin{document}


\title[Multi-primitive IMC for MCTS]{Multi-primitive in-memory computing for Monte Carlo tree search}


\author*[1,2]{\fnm{Tergel} \sur{Molom-Ochir}}\email{tergel.molom-ochir@duke.edu}
\author[1]{\fnm{Benjamin F.} \sur{Morris III}}
\author[1]{\fnm{Yintao} \sur{He}}
\author[2]{\fnm{Archit} \sur{Gajjar}}
\author[2]{\fnm{Giacomo} \sur{Pedretti}}
\author[1]{\fnm{Hai ``Helen''} \sur{Li}}
\author[1]{\fnm{Yiran} \sur{Chen}}
\author[2]{\fnm{Jim} \sur{Ignowski}}
\author[2]{\fnm{Aishwarya} \sur{Natarajan}}
\affil*[1]{\orgdiv{Department of Electrical and Computer Engineering}, \orgname{Duke University}, \orgaddress{\city{Durham}, \postcode{27708}, \state{NC}, \country{USA}}}
\affil[2]{\orgname{Hewlett Packard Labs}, \orgaddress{\city{Milpitas}, \postcode{95035}, \state{CA}, \country{USA}}}


\abstract{
Monte Carlo tree search (MCTS) enables artificial intelligence (AI) decision-making, but requires 55--300\,W on conventional processors, limiting edge deployment. In-memory computing (IMC) is energy-efficient on regular workloads but has been considered incompatible with irregular multi-phase algorithms. We introduce phase-to-primitive decomposition, which reformulates each algorithmic phase as a hardware-native IMC primitive. Applied to MCTS, selection, expansion, rollout and backpropagation map to content-addressable memory, combinational logic, a resistive random-access memory (RRAM) crossbar and static random-access memory, keeping search on chip. At 22\,nm with fabricated RRAM-array parameters, IMC-MCTS consumes $\sim$60\,mW for $9{\times}9$ Go, achieving $96\times$ energy efficiency over a central processing unit (CPU) and $65\times$--$2{,}059\times$ over an H100 graphics processing unit (GPU). It reaches a European Go Federation rating within sample-size uncertainty of open-source Go engines (Pachi-UCT and Michi-C). The same substrate runs eight applications across four AI domains.}


\maketitle

\makeatletter
\global\@topnum=4
\makeatother
\setcounter{topnumber}{4}
\setcounter{bottomnumber}{2}
\setcounter{totalnumber}{6}
\renewcommand{\topfraction}{0.95}
\renewcommand{\bottomfraction}{0.5}
\renewcommand{\textfraction}{0.05}
\renewcommand{\floatpagefraction}{0.7}

\newpage


\begin{figure}[!htbp]
\centering
\begin{subfigure}[t]{0.98\linewidth}
  \centering
  \includegraphics[width=\linewidth,height=4.5cm,keepaspectratio]{Figures/fig01_decomposition/1d_mcts_phases.pdf}
  \caption{}\label{fig:concept_a}
\end{subfigure}\\[0.8em]
\begin{subfigure}[t]{0.48\linewidth}
  \centering
  \includegraphics[width=\linewidth,height=4.5cm,keepaspectratio]{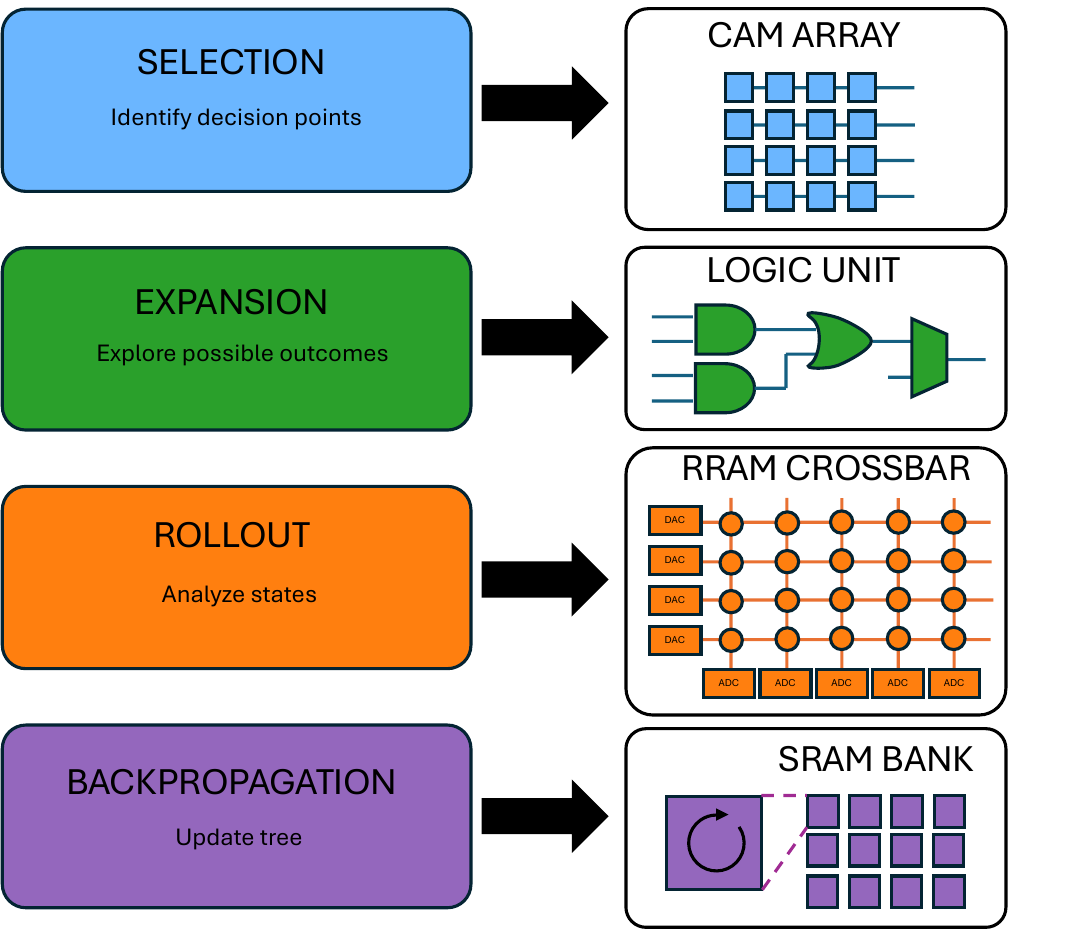}
  \caption{}\label{fig:concept_b}
\end{subfigure}\hfill
\begin{subfigure}[t]{0.48\linewidth}
  \centering
  \includegraphics[width=\linewidth,height=4.5cm,keepaspectratio]{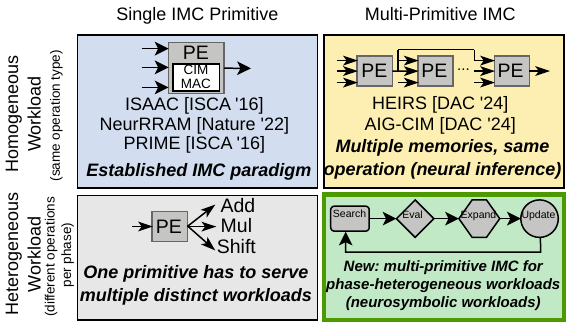}
  \caption{}\label{fig:concept_c}
\end{subfigure}
\caption{\textbf{Phase-to-primitive decomposition.}
\textbf{a}, Monte Carlo tree search (MCTS) iterates through four phases (selection, expansion, rollout, backpropagation) over a search tree.
\textbf{b}, Phase-to-primitive mapping: each phase corresponds to a co-located in-memory computing (IMC) primitive (content-addressable memory (CAM), combinational logic, resistive random-access memory (RRAM) crossbar, static random-access memory (SRAM)).
\textbf{c}, The IMC workload landscape, with prior IMC accelerators occupying the homogeneous quadrants and IMC-MCTS in the phase-heterogeneous, multi-primitive quadrant.}
\label{fig:concept}
\end{figure}

Monte Carlo tree search (MCTS)\cite{coulom2006efficient,uct} underpins artificial intelligence (AI) decision-making in domains from world-champion game-playing agents\cite{alphago,silver2017alphazero} to robotic path planning\cite{dam2022robot}, protein-folding heuristics\cite{deng2022protein} and combinatorial optimization\cite{buzer2023mcts}. Yet its 55--300\,W power draw on conventional processors\cite{intel_xeon_tdp,nvidia_gpu_power} confines it to data-center hardware. This places MCTS out of reach of the milliwatt-scale embedded platforms (drones, surgical robots, battery-powered scientific instruments) where autonomous decision-making under tight energy budgets is increasingly needed. The cause is structural: MCTS's irregular tree traversal, control-flow-heavy expansion and statistical update phases parallelize inefficiently on conventional multi-core and GPU architectures\cite{chaslot2008parallel,rocki2011parallel}. Full neural-guided MCTS systems therefore run on data-center-scale hardware (the distributed AlphaGo used 1{,}202 central processing units (CPUs) and 176 graphics processing units (GPUs))\cite{alphago}.

Closing this gap requires moving computation to where the data lives. In-memory computing (IMC)\cite{ielmini2018memory} does exactly that, co-locating computation with storage and eliminating the data movement that dominates conventional architectures. IMC has produced dramatic energy efficiency gains for regular workloads (deep neural-network inference\cite{wan2022nature}, kernel approximation\cite{sebastian2024kernel}, recurrent networks\cite{li2019lstm} and reinforcement-learning policy networks\cite{portner2025actor}) by mapping a single dominant operation, typically matrix-vector multiplication, onto a single IMC primitive. Algorithms with multiple, qualitatively distinct phases (e.g.\ MCTS's selection, expansion, rollout and backpropagation) have not previously been mapped onto IMC, to our knowledge, because no single primitive can serve all of their operations efficiently.

Here we show that this incompatibility is not fundamental. We introduce \emph{phase-to-primitive decomposition}, a four-step methodology (profile, match, reformulate, compose) that converts a phase-heterogeneous algorithm into a multi-primitive IMC workload. The methodology requires that each algorithmic phase be reformulated to expose a hardware-native operation, after which the phases compose into an end-to-end pipeline that keeps all data on-chip. The composition of multiple, qualitatively distinct primitives within one substrate parallels the heterogeneous, anatomically specialized organization of biological neural systems\cite{marr1971archicortex,treves1994ca3} and aligns with principles in neuromorphic engineering\cite{mead1990neuromorphic,sebastian2020imc}; however, we make no claim of direct correspondence between any specific primitive and any specific brain region. We apply this methodology to MCTS, demonstrating that an algorithm previously regarded as IMC-incompatible can be mapped end-to-end onto co-located memory primitives (Fig.~\ref{fig:concept}).

The resulting accelerator (IMC-MCTS) executes all four MCTS phases on co-located memory primitives: content-addressable memory (CAM)\cite{pagiamtzis2006content} performs $\mathcal{O}(1)$ associative lookup of tree nodes, replacing $\mathcal{O}(n)$ hash-chain traversal; combinational logic generates legal moves in a single-cycle pass, replacing sequential software loops; an analog resistive random-access memory (RRAM) crossbar\cite{hu2018memristor,li_cmos-integrated_2020} performs learned position evaluation in the analog domain at $0.308$\,pJ per multiply--accumulate (MAC) operation; and static random-access memory (SRAM) holds tree statistics that are updated in-place, eliminating off-chip data transfer. A compact finite-state machine (FSM) orchestrates a four-stage pipeline that maintains deterministic timing.

We test IMC-MCTS on the game of Go, chosen because its $10^{170}$ legal positions make it a stringent test of decision-making and a long-standing benchmark for AI\cite{tromp2016number,silver2017alphazero}. In a 1{,}050-game round-robin tournament (50 games per pairing) against six engines ranging from random play to KataGo\cite{wu2019katago}, IMC-MCTS hardware reached a final European Go Federation (EGF) tournament rating of $\approx$1727 (3 kyu) at matched 500-simulation compute, on par with Michi-C ($\approx$1706 EGF) and Pachi-UCT (Pachi using Upper Confidence bounds applied to Trees) without hand-coded priors ($\approx$1660 EGF). This is achieved despite the hardware using only an analog evaluator with an 8-bit analog-to-digital converter (ADC) and binary (1-bit) digital-to-analog converter (DAC) input encoding, with floating-point differential-conductance weights.

To test versatility, we evaluate the same hardware, without architectural modifications, by reprogramming crossbar weights, on eight applications across four AI domains: strategy games (Connect Four, Othello, Hex, Go), navigation (FrozenLake, MiniGrid), scientific optimization (hydrophobic-polar (HP) protein folding) and puzzles (Minesweeper). All eight tasks show observed better play over random rollouts, with Search Guidance Gain (SGG; Methods) ranging from $+0.09$ to $+0.56$. All eight exceed the threshold of $+0.05$ used to mark visible effects in these 10-game demonstrations (Methods; Supplementary Note~9.3).

The energy savings are substantial: energy per move scales from $\sim$3\,$\mu$J at $8{\times}8$ boards to $\sim$18\,$\mu$J at $13{\times}13$. At $9{\times}9$ Go, IMC-MCTS consumes $\sim$60\,mW during sustained 5{,}000-iteration-per-move search (Methods), roughly three orders of magnitude below the 55--300\,W power draw of CPU and GPU MCTS implementations, achieving $96\times$ energy efficiency over CPU and $65\times$--$2{,}059\times$ over an H100 GPU depending on neural-MCTS batch size (single-position to batched-256; Fig.~\ref{fig:energy}c).

These results demonstrate that the apparent incompatibility between IMC and irregular AI algorithms is a property of the operating-point chosen rather than a fundamental limit, and that milliwatt-scale, multi-domain AI decision-making (previously confined to data-center processors) can be supported on edge hardware.

\section*{Results}\label{sec:results}

\subsection*{Phase-to-primitive decomposition}

Phase-to-primitive decomposition treats a heterogeneous algorithm as a sequence of computational phases, each of which is independently mapped onto an IMC primitive matching its dominant operation pattern. The methodology has four steps. First, we \emph{profile} each phase and classify its compute pattern as memory-bound, compute-bound, control-bound or deterministic. Second, we \emph{match} each pattern to an IMC primitive class: associative search to CAM, dense linear algebra to an analog crossbar, sequential read-modify-write to SRAM, and fixed regular logic to combinational hardware. Third, we \emph{reformulate} phases that are incompatible with their matched primitive into compatible operations; this step may require algorithmic changes. Fourth, we \emph{compose} the primitives into an end-to-end pipeline with explicit handshaking, so that no phase reverts to off-chip storage.

Applying the methodology to MCTS exposes four phases with distinct computational patterns: selection (memory-bound, irregular), expansion (control-bound, deterministic), rollout (compute-bound, approximate-tolerant) and backpropagation (memory-bound, sequential). Crucially, the primary bottleneck is highly dependent on the target platform (Extended Data Fig.~\ref{eda:bottlenecks}, Extended Data Table~\ref{edt:phase_breakdown}). On a CPU, rollout dominates execution; whether a branchy random play-out or a dense matrix-vector product, it forms the largest per-iteration block of serial compute. On GPU under traditional MCTS, backpropagation dominates instead, consuming 68.2\% of execution time on the H100 because of per-iteration Compute Unified Device Architecture (CUDA) kernel-launch overhead and counter-update contention in the \texttt{mcts\_numba\_cuda} reference. Neural-MCTS amortizes these updates across larger CUDA Basic Linear Algebra Subprograms (cuBLAS) batches, shrinking backpropagation to a negligible share of GPU time, but the dominant phase still moves. Because the bottleneck shifts across conventional platforms, accelerating any single phase is insufficient: an end-to-end, multi-primitive approach is required to accelerate the full pipeline.

Three reformulations make MCTS fully IMC-compatible. First, we replace variable-length random rollouts with a fixed-latency, two-layer neural evaluator, as crossbars require fixed-dimension inputs. Second, we restructure pointer-chasing tree traversal into a parallel associative lookup enabling CAM. Third, we convert off-chip statistics updates to in-place SRAM writes. Together with the finite state machine as the orchestrator, the resulting four primitives reinforce one another. The CAM's single-cycle latency prevents stalls that would otherwise idle the analog crossbar, while the crossbar's deterministic timing permits a fixed-stage pipeline. This load-bearing property, where each hardware primitive depends on the others' timing guarantees, is what distinguishes our phase-to-primitive decomposition from traditional, isolated modular acceleration.

\subsection*{Architecture and operation}

\begin{figure}[!htbp]
\centering
\includegraphics[width=0.98\linewidth]{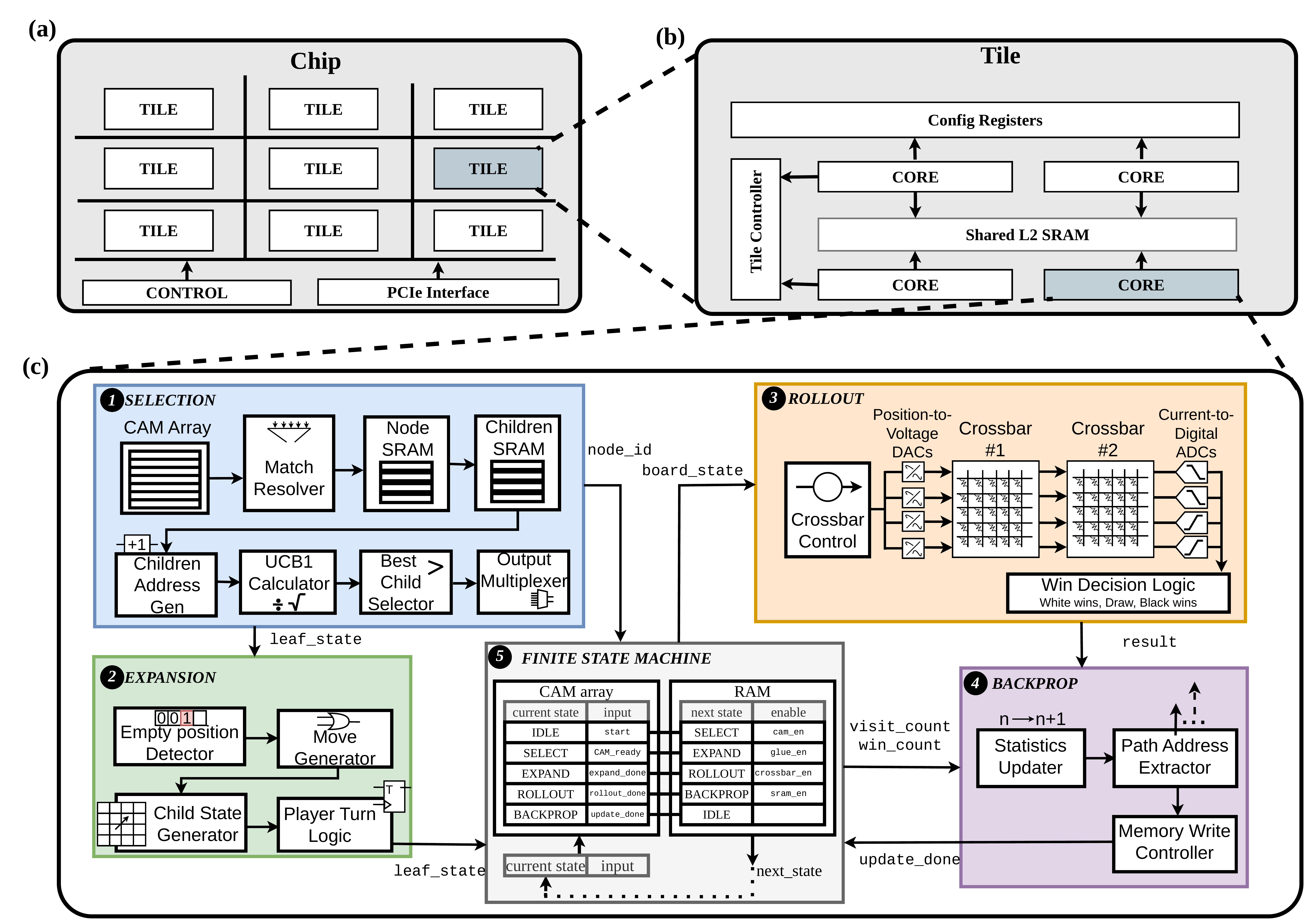}
\\[0.8em]
{\renewcommand{\thesubfigure}{d}%
\begin{subfigure}[t]{0.48\linewidth}
  \centering
  \includegraphics[width=\linewidth,height=4.5cm,keepaspectratio]{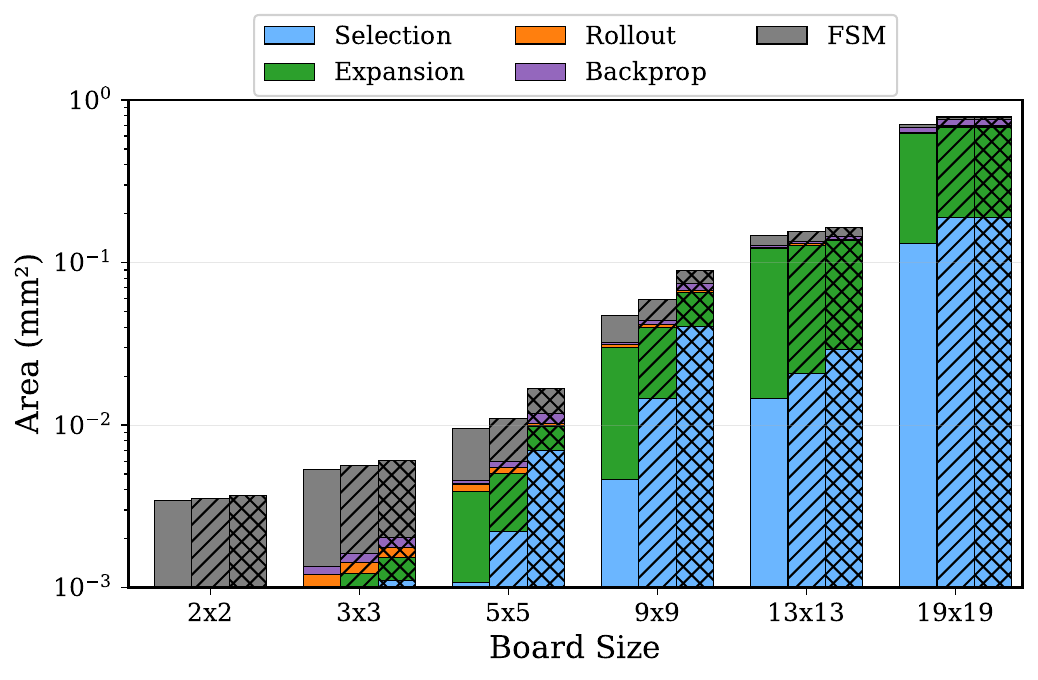}
  \caption{}\label{fig:architecture_d}
\end{subfigure}}\hfill
{\renewcommand{\thesubfigure}{e}%
\begin{subfigure}[t]{0.48\linewidth}
  \centering
  \includegraphics[width=\linewidth,height=4.5cm,keepaspectratio]{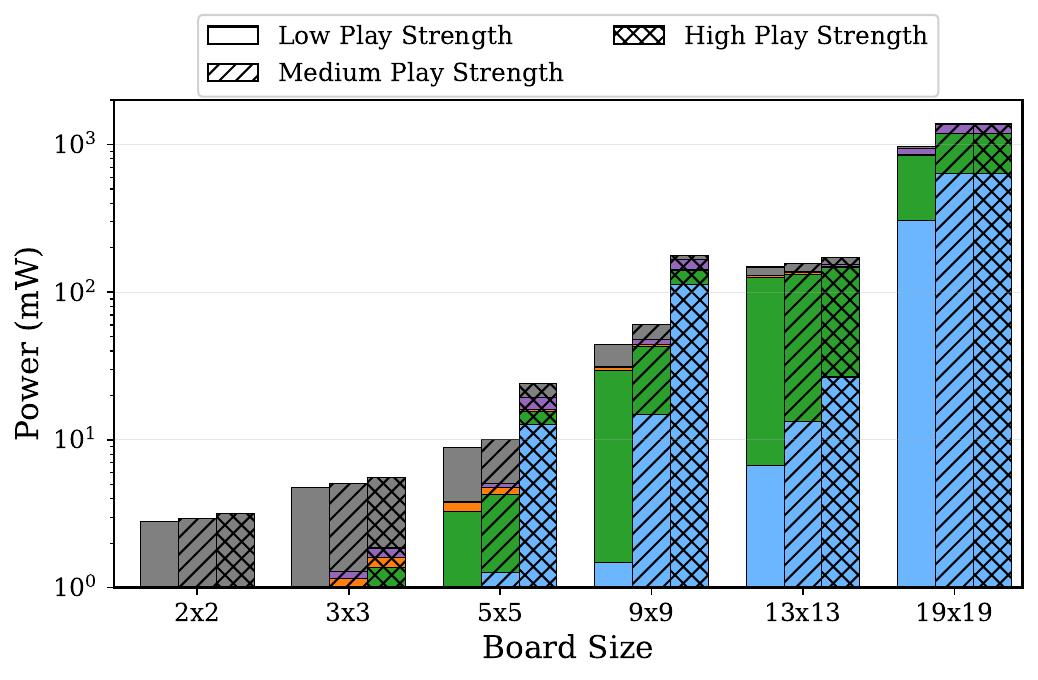}
  \caption{}\label{fig:architecture_e}
\end{subfigure}}
\caption{\textbf{IMC-MCTS architecture and resource scaling.}
\textbf{a}, Chip view: multiple tiles connected through a top-level control block and a host Peripheral Component Interconnect Express (PCIe) interface.
\textbf{b}, Tile view: cores share a level-2 (L2) SRAM through a tile controller.
\textbf{c}, Core schematic: one core implements the four-stage MCTS pipeline (Selection, Expansion, Rollout, Backpropagation) under TCAM-based finite-state-machine control.
\textbf{d}, Per-component area scaling from $2{\times}2$ to $19{\times}19$ Go.
\textbf{e}, Per-component power distribution during sustained 5{,}000-iteration-per-move search at $9{\times}9$, summing to 60.26\,mW. Per-component values are in Supplementary Table~S3.}
\label{fig:architecture}
\end{figure}

The IMC-MCTS accelerator is organized as a tiled chip with a host Peripheral Component Interconnect Express (PCIe) interface (Fig.~\ref{fig:architecture}a). Each tile contains multiple cores sharing a level-2 (L2) SRAM (Fig.~\ref{fig:architecture}b), and each core implements the full four-stage MCTS pipeline (Fig.~\ref{fig:architecture}c). The selection stage stores tree-node board states as bit patterns in a six-transistor (6T) SRAM-based CAM and reads Upper Confidence Bound (UCB1) statistics from a tightly coupled SRAM hierarchy; an issued query returns the addresses of all child nodes in parallel, and a tournament selector identifies the best child within a single-cycle decision. The expansion stage uses a Game Logic Unit (GLU) that evaluates all $N^2$ board cells simultaneously through combinational logic, producing legal-move successor states deterministically. The rollout stage encodes the leaf state through binary (1-bit) DACs and performs analog matrix-vector multiplication on RRAM crossbars storing pre-trained differential-conductance weights ($162{\times}96$ and $96{\times}3$ for $9{\times}9$ Go). 8-bit ADCs then digitize the output, and a softmax yields win/loss/draw probabilities. The backpropagation stage performs in-place read-modify-write on SRAM along the leaf-to-root path, using saturating counters to prevent overflow. A finite-state machine implemented with a ternary content-addressable memory (TCAM) orchestrates these four stages with deterministic timing.

The four-stage pipeline matches the four MCTS phases. Like other resistive in-memory compute accelerators\cite{wan2022nature}, the crossbar uses an 8-bit ADC datapath against binary (1-bit) DAC input encoding (the standard operating point for analog in-memory neural-network accelerators\cite{shafiee2016isaac}) and stores weights as differential-conductance pairs. Methodological details (synthesis flow, register-transfer level (RTL) hierarchy, SRAM characterization via the CACTI 6.0 memory modeling tool\cite{cacti}, cycle-accurate simulation through the Structural Simulation Toolkit (SST)\cite{sst}, and crossbar noise-robustness analysis) are reported in Methods and Supplementary Notes 3--5.

\subsection*{Decision quality on Go}

We evaluated the playing strength of IMC-MCTS hardware on $9{\times}9$ Go through a 1{,}050-game round-robin tournament against six reference engines: KataGo\cite{wu2019katago} (strongest reference engine), Pachi-UCT\cite{baudis2011pachi}, Michi-C\cite{baudis_michi}, GnuGo at level 10, an MCTS-trained-with-supervision baseline (IMC-weak), and a uniform-random baseline. Tournament Elo ($K=32$, initial 1500) is calibrated against GnuGo-L10's standard EGF rating of 1500 (6 kyu). KataGo settles at $\approx$3200 EGF in the tournament, consistent with its commonly reported strong-amateur range. All MCTS players used 500 simulations per move under standard Go tournament rules (full configuration in Methods).


The IMC-MCTS hardware reaches an EGF rating within sample-size uncertainty of two established open-source MCTS engines, Pachi-UCT and Michi-C, in the 1{,}050-game round-robin tournament at 500 simulations per move (Fig.~\ref{fig:energy}d). Pachi's iso-compute configuration is detailed in Methods. We evaluate two evaluator-training regimes on identical hardware. \emph{IMC-strong} uses a self-play-trained evaluator, iteratively refined inside the IMC-MCTS search loop over 10 iterations of $\sim$35{,}000 positions each, sparring against Pachi-UCT, and reaches $\sim$96\% three-class validation accuracy. \emph{IMC-weak} uses the same supervised training pipeline but is deliberately checkpointed at 60\% three-class validation accuracy, providing a low-quality-evaluator baseline on identical silicon (Methods).

Final placements on the tournament EGF scale were as follows: IMC-strong $\approx$1727 (3 kyu), Michi-C $\approx$1706 (3 kyu), Pachi-UCT no-patterns $\approx$1660 (4 kyu), GnuGo-L10 $\approx$1500 (6 kyu, calibration anchor), KataGo $\approx$3200 (upper reference), IMC-weak $\approx$1139 (9 kyu). The pairwise gaps separating IMC-strong from Michi-C ($\Delta\approx$21 Elo) and from Pachi-UCT ($\Delta\approx$67 Elo) are small at the 50-games-per-pairing sample size used here. We therefore report the relative ordering and defer formal separation claims pending the larger-$n$ validation planned for follow-on work (Supplementary Note~9.3). No decision-quality degradation relative to software MCTS was observed within the sample size used at matched compute. The averaging across hundreds of rollouts per move stabilizes decision quality against the analog substrate's noise. Individual position evaluations may be perturbed by the modeled $\sigma\in\{2,5,10\}\%$ conductance noise, but the search ensemble's relative ordering of moves remains robust (Supplementary Note~5).

The IMC-weak comparison isolates evaluator quality from substrate noise: on identical hardware and matched 500-simulation compute, IMC-weak settles at $\approx$1139 EGF (9 kyu), $\sim$590 EGF below IMC-strong, with a head-to-head record of 9--41 (18\% wins). The ${\sim}36$-percentage-point gap in evaluator validation accuracy ($\sim$96\% vs 60\%) tracks the Elo gap. IMC-strong's self-play loop generates training data drawn from the positions encountered at deployment, while IMC-weak does not. This distributional alignment is what mediates the gap: training-data quality, not analog computation, sets playing strength.


\subsection*{Energy efficiency and edge deployment}

Beyond decision quality, the architecture's value rests on its energy profile relative to conventional platforms. We measured IMC-MCTS energy and latency at 22-nm complementary metal-oxide-semiconductor (CMOS) technology against three baseline platforms running matched neural-guided MCTS: Intel Xeon Platinum 8462Y+, AMD Threadripper PRO 5945WX, and NVIDIA H100 (80\,GB High Bandwidth Memory 3 (HBM3)). The H100 was run in both single-position and batched (64- and 256-position) configurations. CPU power was instrumented through Intel Running Average Power Limit\cite{rapl}, GPU power through nvidia-smi.

\begin{figure}[!htbp]
\centering
\begin{subfigure}[t]{0.48\linewidth}
  \centering
  \includegraphics[width=\linewidth,height=5cm,keepaspectratio]{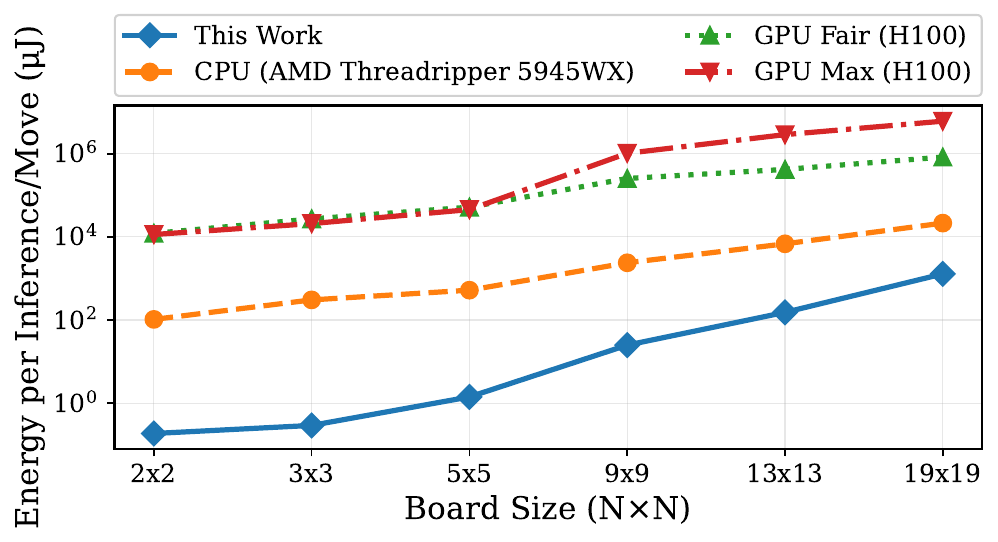}
  \caption{}\label{fig:energy_a}
\end{subfigure}\hfill
\begin{subfigure}[t]{0.48\linewidth}
  \centering
  \includegraphics[width=0.80\linewidth,height=5cm,keepaspectratio]{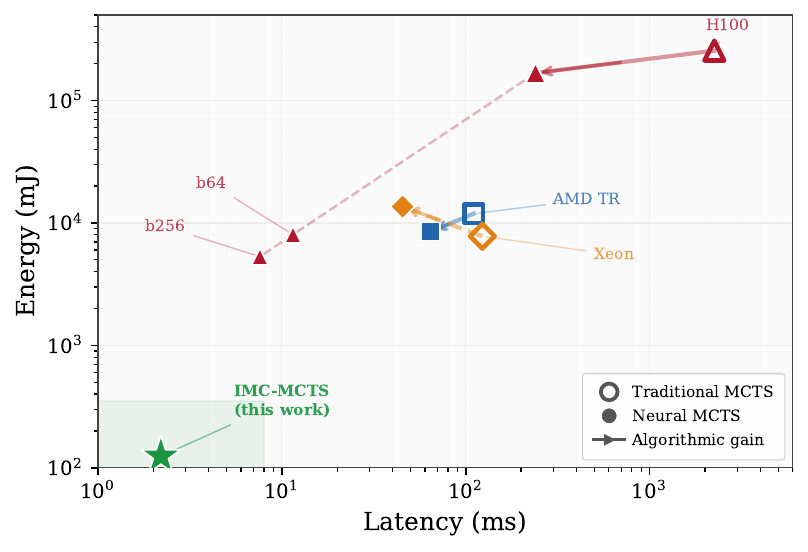}
  \caption{}\label{fig:energy_b}
\end{subfigure}\\[0.8em]
\begin{subfigure}[t]{0.48\linewidth}
  \centering
  \includegraphics[width=\linewidth,height=5cm,keepaspectratio]{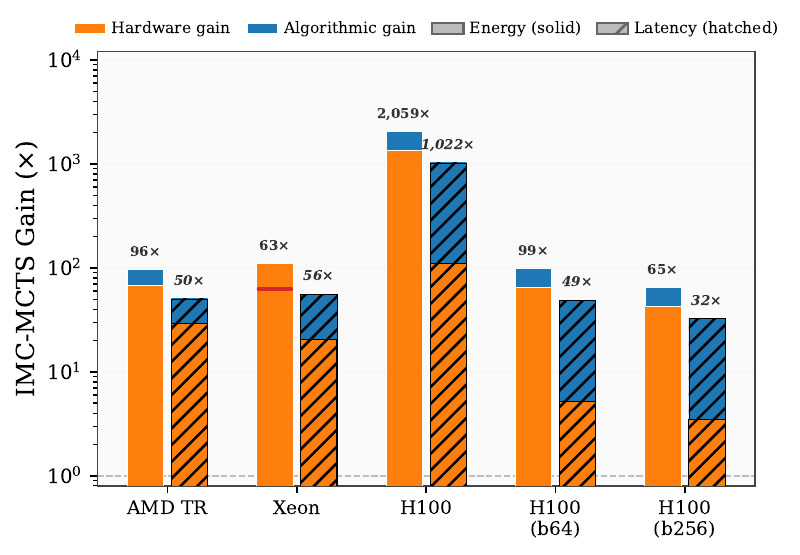}
  \caption{}\label{fig:energy_c}
\end{subfigure}\hfill
\begin{subfigure}[t]{0.48\linewidth}
  \centering
  \includegraphics[width=\linewidth,height=5cm,keepaspectratio]{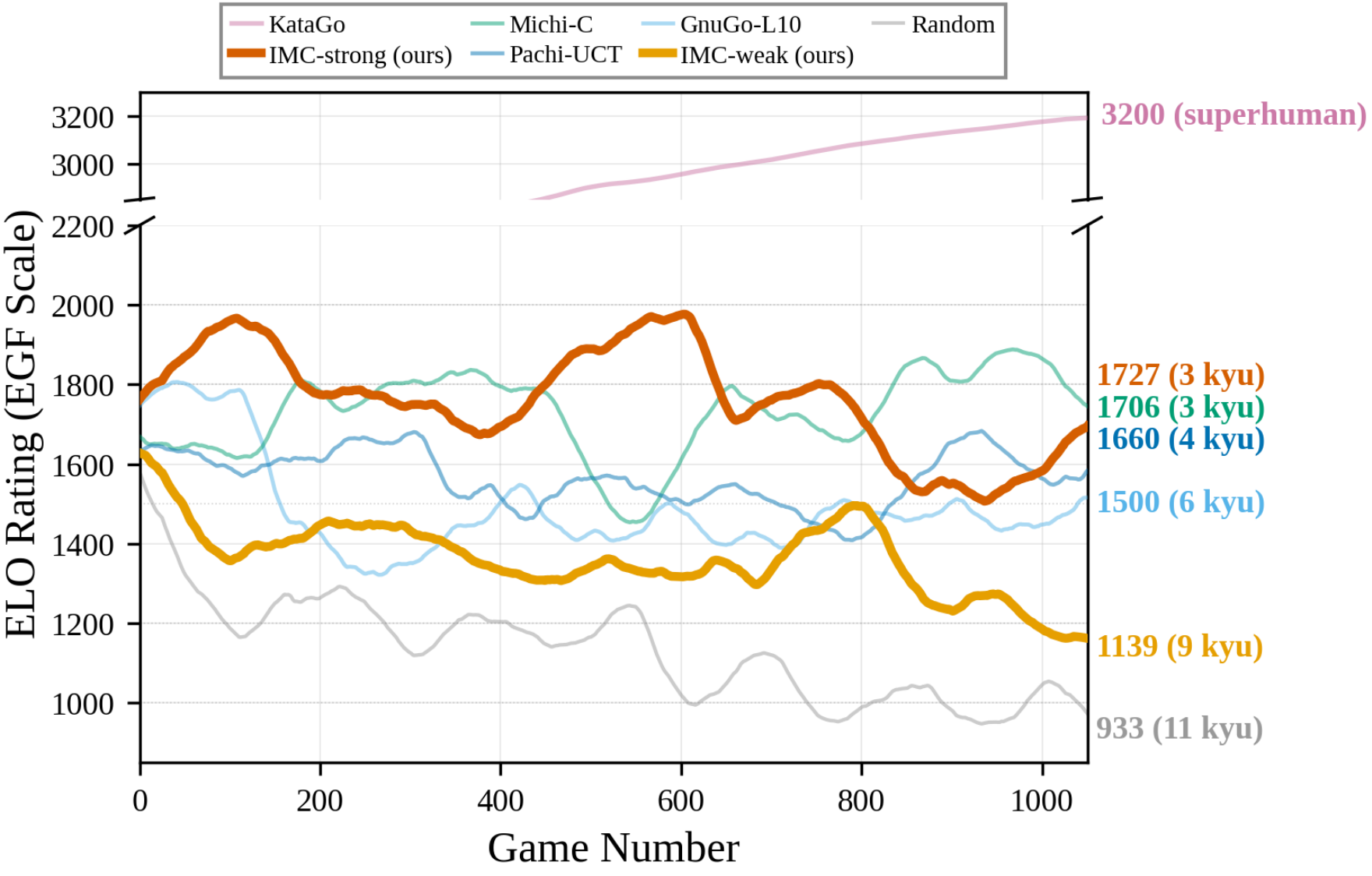}
  \caption{}\label{fig:energy_d}
\end{subfigure}
\caption{\textbf{Energy efficiency and decision quality.}
\textbf{a}, Energy per 5{,}000-iteration MCTS search vs.\ board size, for IMC-MCTS and traditional CPU/GPU MCTS baselines (single-position GPU-Fair: 1 tree, 1 playout; batched GPU-Max: 8 trees, 128 playouts). Neural-MCTS batch-size comparisons appear in panel c.
\textbf{b}, Energy vs.\ latency at $9{\times}9$ Go (Pareto view).
\textbf{c}, Stepwise total gain attribution across five baselines (AMD Threadripper, Intel Xeon, NVIDIA H100 single-position, H100 batch-64, H100 batch-256): algorithmic step (swap to neural evaluator) followed by architectural step (swap to end-to-end IMC), measured as sequential before/after ratios rather than parallel ablations. For AMD Threadripper energy, the blue algorithmic segment is $\approx1.4\times$ and the orange hardware segment is $\approx68\times$, yielding the labeled $96\times$ total.
\textbf{d}, Elo progression during a 1{,}050-game round-robin tournament on $9{\times}9$ Go among seven engines (KataGo, Pachi-UCT, Michi-C, GnuGo-L10, IMC-strong, IMC-weak, uniform-random). Tournament Elo ($K=32$, initial 1500) is calibrated against GnuGo-L10's standard EGF rating of 1500 (6 kyu); final placements are annotated on the right edge. At the 50-games-per-pairing sample size used here, small pairwise gaps (e.g.\ IMC-strong vs Pachi-UCT, $\Delta\approx$67 Elo; IMC-strong vs Michi-C, $\Delta\approx$21 Elo) should be read as ordered rather than statistically separated; larger-$n$ validation is planned for follow-on work.}
\label{fig:energy}
\label{fig:decision_quality}
\end{figure}

At $9{\times}9$ Go, IMC-MCTS delivers $96\times$ energy efficiency over the CPU baseline and $65\times$--$2{,}059\times$ over an H100 GPU depending on neural-MCTS batch size (Fig.~\ref{fig:energy}c). The $65\times$ endpoint is against batched-256 H100 (256-leaf virtual-loss batching, the standard high-throughput operating point). The $2{,}059\times$ endpoint is against single-position H100. Sustained chip power is $\sim$60\,mW during 5{,}000-iteration-per-move search. The batched configuration is the fair operating-point comparison because the single-position figure conflates IMC's architectural advantage with GPU underutilisation (Supplementary Note~6.4). Extended Data Table~\ref{edt:scaling} reports the neural-MCTS hardware-only energy baselines across board sizes. The CPU-baseline gain decomposes stepwise (Fig.~\ref{fig:energy}c): swapping random rollout for the neural evaluator on CPU yields the blue ${\approx}1.4\times$ algorithmic segment, then swapping CPU for IMC at neural-MCTS yields the orange ${\approx}68\times$ hardware segment, multiplying to the $96\times$ headline. This stepwise attribution uses two measured before/after ratios, not parallel ablations. The architectural advantage (constant-time CAM lookup, weight-stationary analog rollout, in-place SRAM updates) outweighs process scaling for irregular workloads.

Throughput ranges from 18{,}471 iterations per second at $2{\times}2$ to 510 at $19{\times}19$. At $9{\times}9$ Go the sustained throughput is ${\approx}2{,}400$ iterations per second. This rate is set by MCTS's strict sequential dependency between iterations (each iteration's selection phase reads tree statistics updated by the previous iteration's backpropagation), not by single-iteration pipeline latency. Despite a 22-nm node against the H100's 4-nm process, IMC-MCTS retains a $3.4\times$ latency advantage over the batched H100 baseline. An iso-process Stillmaker\cite{STILLMAKER201774} projection is provided in Supplementary Note~3.5. Sustained power scales with board dimensions, from $\sim$60\,mW at $9{\times}9$ to $\sim$1.4\,W at $19{\times}19$ (Fig.~\ref{fig:architecture}e), remaining 1--2 orders of magnitude below CPU/GPU operating points across the full scaling range.

These operating points place IMC-MCTS in a power class previously unreachable by CPU or GPU implementations (55--300\,W), opening three new deployment regimes (Supplementary Note~10). \emph{Drone autonomy} operates with sub-10-W total compute budgets; a 55-W planner alone would consume the full budget. \emph{Microcontroller unit (MCU)-class embedded controllers} (sub-100\,mW total power) can run IMC-MCTS as an on-device planning co-processor for motion planning\cite{dam2022robot} and maneuver selection\cite{lenz2016driving}. \emph{Implantable or wearable devices} are reachable through low duty-cycle operation between infrequent inferences. AI decision-making previously confined to data-center processors becomes deployable on coin-cell-powered embedded systems. A detailed comparison against prior MCTS hardware accelerators (field-programmable gate array (FPGA) and GPU) is provided in Supplementary Note~8.

\subsection*{Substrate reusability across grid-based decision tasks}

To demonstrate that the architecture is not Go-specific, we ran the same hardware on eight grid-based decision tasks spanning four AI domains: strategy games (Connect Four, Othello, Hex, Go), grid navigation (FrozenLake, MiniGrid), scientific optimization (HP protein folding) and puzzles (Minesweeper). The pipeline is fully application-agnostic across all five units (CAM, GLU, RRAM crossbar, SRAM, FSM) for the eight applications shown here. The GLU implements an Empty Position Detector that emits a superset of legal moves. The trained crossbar evaluator then suppresses illegal moves at evaluation time, removing the need for game-specific legality logic in hardware (Extended Data Table~\ref{edt:generalisation}, footnote; Supplementary Note~9). The GLU itself accounts for $\sim$1.4\% of total chip area with memory arrays included, or $\sim$43\% of digital logic alone (Supplementary Table~S3). Only the crossbar weights are reprogrammed per task. Per-application hardware customization is a synthesis-time parameter change on a single SystemVerilog design, parameterized by board size and encoding channels (Methods). No RTL is re-written per application.
\begin{figure}[!htbp]
\centering
\begin{subfigure}[t]{0.48\linewidth}
  \centering
  \includegraphics[width=\linewidth,height=5cm,keepaspectratio]{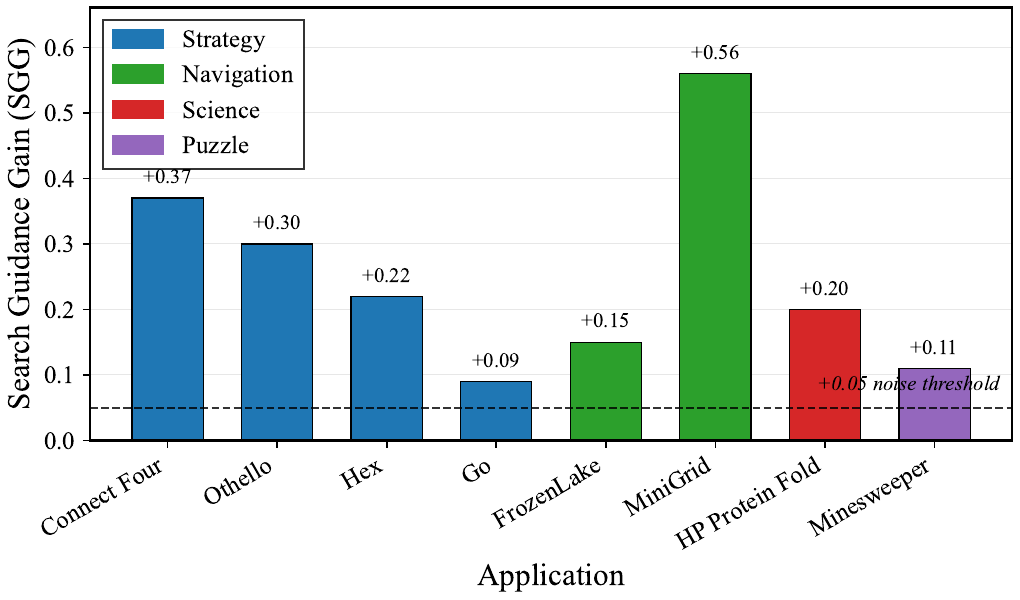}
  \caption{}\label{fig:gen_a}
\end{subfigure}\hfill
\begin{subfigure}[t]{0.48\linewidth}
  \centering
  \includegraphics[width=\linewidth,height=5cm,keepaspectratio]{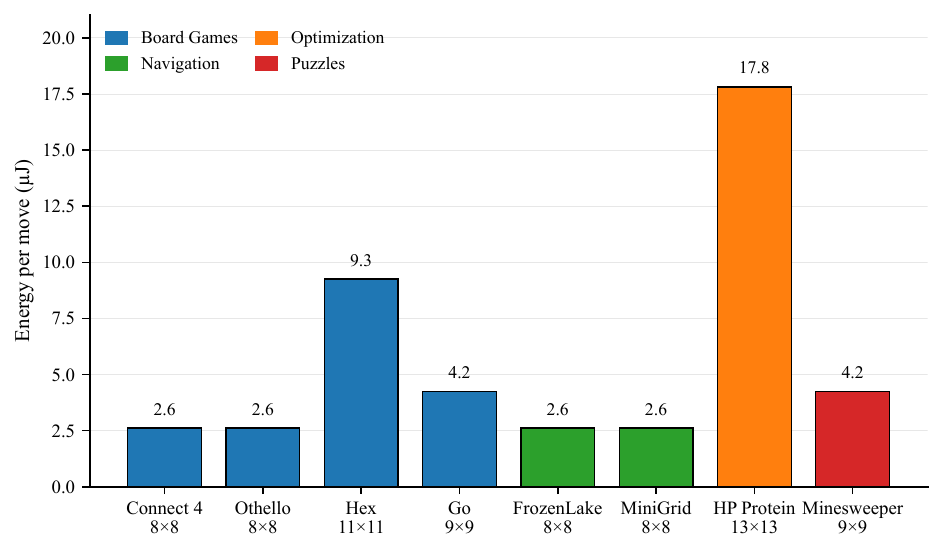}
  \caption{}\label{fig:gen_b}
\end{subfigure}
\caption{\textbf{Substrate reusability across eight grid-based decision tasks.} The same substrate and pipeline execute all eight applications (Connect Four, Othello, Hex, Go, FrozenLake, MiniGrid, HP protein folding, Minesweeper); only the crossbar weights are reprogrammed per task, while CAM, GLU ($\sim$1.4\% of total chip area), SRAM, FSM and crossbar peripherals are unchanged.
\textbf{a}, Search Guidance Gain (SGG; Methods) across the eight applications. The dashed $+0.05$ line is an indicative policy-lift threshold used to mark visible effects in the 10-game demonstrations; it is not a statistical-significance boundary. Specifically, the Go ($+0.09$), Minesweeper ($+0.11$) and FrozenLake ($+0.15$) bars should be read as directional rather than statistically resolved.
\textbf{b}, Per-move energy across the eight applications on identical hardware.}
\label{fig:generalisation}
\end{figure}

All eight applications run on the same hardware with observed positive Search Guidance Gain (SGG) over a random-rollout MCTS baseline (Fig.~\ref{fig:generalisation}a, Extended Data Table~\ref{edt:generalisation}). Intuitively, SGG is the normalized head-to-head win-rate margin between neural-guided MCTS and random-rollout MCTS at matched iteration count (or score margin for single-player tasks). On the reference configuration of 50 iterations and $8{\times}8$ board, SGG~$=0$ means a 50\% win rate (no improvement), SGG~$=+0.2$ corresponds to ${\sim}60\%$, and SGG~$=+0.5$ to ${\sim}75\%$. The normalization makes values comparable across boards and iteration budgets of different size (Supplementary Note~9). Across the eight applications, SGG ranges from $+0.09$ (Go) to $+0.56$ (MiniGrid), all above the $+0.05$ indicative policy-lift threshold used in Fig.~\ref{fig:generalisation}a. The smaller lifts (Go, Minesweeper, FrozenLake) should be read as directional pending larger-$n$ validation. We therefore present this section as a substrate-reusability demonstration rather than a statistical claim of generality. Per-move energy ranges from 2.6 to 17.8\,$\mu$J across applications on identical hardware (Fig.~\ref{fig:generalisation}b). The variation is driven primarily by crossbar size, with $8{\times}8$ boards consuming the least and $13{\times}13$ the most.

\section*{Discussion}\label{sec:discussion}

IMC has so far been treated as a compute methodology for regular workloads (convolution, matrix multiplication, attention) where a single dominant operation can be mapped onto a single primitive\cite{wan2022nature,sebastian2024kernel,fujiwara2022isscc}. The implicit assumption has been that algorithms with multiple computational modes either require general-purpose hardware or revert to off-chip storage between phases, negating IMC's energy advantage. Our results challenge that assumption. Phase-to-primitive decomposition shows that the relevant property is not the regularity of an algorithm, but whether each of its phases admits a hardware-native reformulation. When that condition holds, the phases can be composed into an end-to-end pipeline that retains IMC's data-locality benefits across all phases.

A survey of 12 prior MCTS hardware acceleration efforts (Supplementary Note~8, Supplementary Table~S6; representative anchors in Extended Data Table~\ref{edt:prior_work}), spanning implementations across four FPGAs, six GPUs, one neuromorphic platform, and one multi-core CPU, finds three structural gaps: none uses in-memory computing for any phase, 11 of 12 retain random-simulation rollouts rather than modern neural evaluators, and only one of 12 reports a complete energy-efficiency figure. IMC-MCTS is the first MCTS accelerator to co-deploy multiple in-memory primitives across all four phases of a neural-guided MCTS pipeline, and the first to report end-to-end energy efficiency against both CPU and modern batched-GPU baselines.

For MCTS specifically, the consequence is a $\sim$60\,mW operating point at $9{\times}9$ Go (sustained Medium-strength power), a regime $\sim$$10^3\times$ below the power draw of CPU and GPU MCTS implementations. This shifts the deployable footprint of MCTS from data-center racks to battery-powered or duty-cycled embedded systems, a transition relevant to autonomous robotics, surgical planning, in-the-field scientific optimization and any AI agent that must decide under tight energy budgets. The same substrate executes eight diverse grid-based decision tasks without architectural modifications, demonstrating that the design is not Go-specific. Whether the phase-to-primitive methodology generalizes to non-grid algorithms is a question for future work.

Several limitations bear noting. First, the accelerator targets grid-representable state spaces; non-grid states (continuous robotics, large-graph routing) require encoding modifications, although the four-stage pipeline itself remains applicable. Second, our energy and latency numbers derive from synthesis at 22\,nm with crossbar device parameters drawn from fabricated memristor crossbars\cite{hu2018memristor}, including a 180-nm $64{\times}64$ array\cite{li_cmos-integrated_2020}. While we empirically establish device-non-ideality tolerance up to $\sigma=10$\% additive conductance noise via noise injection through the analog evaluator (Methods; Supplementary Note~5), end-to-end silicon validation of the CAM--crossbar interaction at full pipeline rate is the natural next step. Finally, the analog evaluator's strong-kyu Go play (8-bit ADC datapath, differential-conductance weights) reflects the quality of self-play training data more than the precision of the analog substrate. This is highly favorable for adoption, as algorithmic improvements in training transfer directly to the hardware. However, it means stronger play (e.g.\ professional dan level) requires both higher-quality training data and crossbar capacity scaling.

Beyond MCTS, phase-to-primitive decomposition offers a methodology for accelerating other control-flow-heavy AI algorithms whose phases admit primitive-compatible reformulations: constraint satisfaction\cite{cai2020optimization}, planning under partial observability, and neuro-symbolic reasoning. As AI agents increasingly couple neural perception with symbolic decision-making, the demand for milliwatt-scale decision hardware will move from a luxury to a necessity. Our results indicate that such hardware is now within reach. The path to it lies through algorithmic decomposition rather than through ever-larger general-purpose accelerators.


\section*{Methods}\label{sec:methods}

\subsection*{Phase-to-primitive decomposition (overview)}

The four-step methodology (profile, match, reformulate, compose) is described in full in Supplementary Note 2. In brief, profiling identifies each phase's compute pattern (memory-bound, compute-bound, control-bound, deterministic); matching pairs each pattern to an IMC primitive class (associative search to CAM, dense linear algebra to analog crossbar, sequential read-modify-write to SRAM, fixed regular logic to combinational hardware); reformulation rewrites incompatible phases into primitive-compatible operations (Supplementary Note 2.3 details the three reformulations required for MCTS); and composition wires the primitives into an end-to-end pipeline with explicit handshaking. The decomposition is load-bearing: ablating any single primitive substantially degrades end-to-end performance, with per-component multipliers (energy and latency itemized separately) detailed in Supplementary Note~7.

\subsection*{IMC-MCTS architecture and synthesis}

The accelerator is implemented as a hierarchical SystemVerilog design supporting board sizes from $2{\times}2$ to $19{\times}19$, with conditional generate blocks synthesizing only the logic required for the chosen board size. Five hardware units (CAM-based selection, combinational expansion via a Game Logic Unit, analog RRAM crossbar rollout, in-memory SRAM backpropagation, and a TCAM-based finite-state-machine controller) are connected through a custom interconnect with valid/ready handshakes. Digital components were synthesized with Synopsys Design Compiler against a Taiwan Semiconductor Manufacturing Company (TSMC) 65-nm standard-cell library at 500\,MHz, typical-typical corner, $1.2$\,V, and scaled to 22\,nm via Stillmaker's technology-independent scaling\cite{STILLMAKER201774}. SRAMs were characterized through the CACTI 6.0 memory modeling tool\cite{cacti}; the CAM\cite{pagiamtzis2006content,auth2012cmos} and a conventional 16T TCAM\cite{pagiamtzis2006content} uses prior implementations scaled to 22\,nm. Per-component area and power breakdowns at $9{\times}9$ Go appear in Fig.~\ref{fig:architecture}d, e; full RTL hierarchy, verification methodology and synthesis parameters appear in Supplementary Notes 3 and 4.

\subsection*{Crossbar device modeling}

The analog rollout unit uses RRAM crossbars in a 1T1R (one-transistor, one-memristor) configuration\cite{sheng2019low,hu2018memristor,li_cmos-integrated_2020}, with differential-conductance weight encoding in the 0.1--99.9\,$\mu$S range. The ADC datapath is 8-bit (the standard operating point for analog in-memory neural-network accelerators) and the input encoding is binary (1-bit DAC: $V_\text{high}=1.0$\,V, $V_\text{low}=0$\,V). Crossbar dimensions scale with board size: $9{\times}9$ Go uses two arrays ($162{\times}96$ and $96{\times}3$); for $19{\times}19$ ($722{\times}128$), the first layer partitions across three sub-arrays ($\le$256 rows each), preserving single-cycle semantics. The 768-ns read latency decomposes into 256\,ns of wordline charging and 512\,ns of ADC conversion, characterized from a fabricated 180-nm CMOS $64{\times}64$ RRAM crossbar with complete ADC and DAC peripherals\cite{li_cmos-integrated_2020} and scaled to 22\,nm.

Crossbar weights derive from two-layer feedforward networks scaling with board size (8--722 inputs, 16--192 hidden neurons, 3-class softmax output), trained with AdamW ($\eta = 0.003$) on 2{,}000 labeled positions per board size with early stopping. Three-class accuracy varies with task complexity: 60\% on $9{\times}9$ Go (the largest and noisiest action space), 80\% on Connect Four, Othello and Hex, 93\% on Minesweeper, and 100\% on FrozenLake, MiniGrid and HP protein folding (per-application values in Extended Data Table~\ref{edt:generalisation}). Per-application accuracy reflects substrate demonstration rather than policy optimization: the 2{,}000-position training budget per board size is intentionally light, so that any observed playing strength is attributable to the hardware--algorithm composition rather than to extensively-trained networks. Scaling the training set is orthogonal to the architectural contribution of this paper; larger training corpora are expected to improve absolute playing strength but would obscure the substrate-vs-data attribution that the present comparison makes.

Crossbar non-idealities (device-to-device conductance variation, programming write noise, drift) were modeled by injecting additive Gaussian noise of $\sigma \in \{2, 5, 10\}$\% (matching reported RRAM programming-variation envelopes\cite{hu2018memristor,li_cmos-integrated_2020,wan2022nature}) at each forward pass. End-to-end MCTS robustness across this noise range (quantified through win-rate against a matched random-rollout baseline at each $\sigma$, together with per-$\sigma$ 1-ply policy agreement) is reported in Supplementary Note~5.

\subsection*{Cycle-accurate simulation}

A custom discrete-event simulator built on the Structural Simulation Toolkit (SST)\cite{sst} models cycle-accurate timing across the full pipeline. Each pipeline unit is implemented as an SST component communicating via 1--5\,ns delay links. We simulate 18 configurations (six board sizes from $2{\times}2$ to $19{\times}19$ $\times$ three play-strength levels: low, medium, high; per-configuration iteration budgets in Supplementary Table~S1). Functional verification used SystemVerilog testbenches with directed and constrained-random stimuli, complemented by differential testing against a cycle-accurate Python golden model across more than 1{,}000 randomized games to confirm functional equivalence and end-to-end throughput.

\subsection*{Iteration budgets and energy metrics}

Different results in this paper use different MCTS iteration budgets per move, summarized here for clarity. The Go decision-quality tournament used \textbf{500 simulations per move} (iso-compute comparison against software MCTS engines). The sustained energy and power measurements, including the 60.26\,mW headline at $9{\times}9$ and the energies in Extended Data Table~2, used \textbf{5{,}000 iterations per move} (Medium strength). High-strength scaling experiments used up to \textbf{50{,}000 iterations per move} (Supplementary Table~S1). Cross-domain applications (Extended Data Table~3, Fig.~4) use per-task budgets in the 50--200 iterations-per-move range, also tabulated in Supplementary Table~S1.

Energy is reported throughout at the level of a full $N$-iteration move (energy per move $\equiv$ energy per $N$-iteration search at a given $N$); sustained power is the chip's operating power during search; energy-efficiency gain is the ratio of baseline energy per move to IMC energy per move under matched $N$. Units (mJ vs $\mu$J) reflect the absolute magnitude at the configuration in question: Go at Medium strength (5{,}000 iter/move) reports mJ per move (Extended Data Table~2), while cross-domain applications at 50--200 iter/move report $\mu$J per move (Fig.~4b, Extended Data Table~3).

\subsection*{Tournament evaluation}

We validated decision quality through comprehensive round-robin tournaments with sample-size-aware interpretation. The evaluation used a 1{,}050-game round-robin tournament on $9{\times}9$ Go with seven engines: KataGo\cite{wu2019katago}, Pachi-UCT\cite{baudis2011pachi}, Michi-C\cite{baudis_michi}, GnuGo at level 10, IMC-strong (self-play-trained evaluator, $\sim$96\% three-class validation accuracy on held-out self-play positions; 10 iterations of $\sim$35{,}000 positions each, sparring against Pachi-UCT), IMC-weak (the same supervised training pipeline deliberately checkpointed at 60\% three-class validation accuracy to provide a low-quality-evaluator baseline on identical silicon; per-application accuracies for the separate cross-domain demonstration networks are tabulated in Extended Data Table~\ref{edt:generalisation}), and a uniform-random baseline. Each of the 21 matchups comprised 50 games with alternating colours. All MCTS players used 500 simulations per move; KataGo used its default playout schedule. Pachi-UCT was run with $3{\times}3$ pattern libraries disabled to match the iso-compute with no-prior-knowledge configuration of the other MCTS players. This is below Pachi's competitive operating point (10k--100k playouts with patterns enabled), but it is the necessary configuration for a fair head-to-head comparison at matched compute.

Games were scored under Chinese area scoring with a 6.5-point komi. Tournament Elo ratings ($K=32$, initial 1500) are calibrated against GnuGo-L10's standard EGF rating of 1500 (6 kyu); KataGo's tournament rating of $\approx$3200 EGF is consistent with its commonly reported strong-amateur range and serves as the upper reference. Pachi-UCT no-patterns ($\approx$1660 EGF, 4 kyu) and Michi-C ($\approx$1706 EGF, 3 kyu) settle to their natural tournament placements reflecting head-to-head results. The relative ordering of engines is set by win rates and is invariant to the choice of upper-anchor pinning. All MCTS players used UCB1 selection (Eq.~1, Supplementary Note~1.2); the predictor-UCB applied to trees (PUCT) extension\cite{silver2017alphazero} is supported by the hardware (Supplementary Note~1.3) but is not exercised in our experiments, which keep selection comparable across all tested engines.\subsection*{Baselines and energy measurement}

CPU baselines (Intel Xeon Platinum 8462Y+, AMD Threadripper PRO 5945WX) used optimized C++17 (\texttt{-O3}) with single-threaded Upper Confidence bounds applied to Trees (UCT)\cite{uct}; GPU baselines (NVIDIA H100) used both traditional MCTS via \texttt{mcts\_numba\_cuda}\cite{mctsnc_klesk2025} and neural MCTS via the Compute Unified Device Architecture (CUDA) Basic Linear Algebra Subprograms (cuBLAS) library, in single-position and batched (64-, 128-, 256-position) configurations. CPU power was measured via Intel Running Average Power Limit\cite{rapl}; GPU power via \texttt{nvidia-smi}. All baselines use single-position inference per MCTS iteration (matching the algorithm's sequential dependency); batched configurations are reported separately to ensure fair comparison against optimized GPU operating points. Both platforms implement random-rollout and neural-network variants across six board sizes.

\subsection*{Cross-domain evaluation protocol}

For each of the eight applications (Connect Four, Othello, Hex, Go, FrozenLake, MiniGrid, HP protein folding, Minesweeper), per-application neural networks were trained on self-play data with task-specific input encodings (Extended Data Table~\ref{edt:generalisation}). Hardware area and energy were computed using the parameterized synthesis model described in Supplementary Note 3; the same pipeline processes all eight applications without architectural modifications, with only crossbar weights reprogrammed per application. SGG values were averaged over 10-game matchups against random-rollout MCTS on the same board size and iteration budget; the $+0.05$ indicative cutoff and its 10-game sample-size limitations are detailed in Supplementary Note~9.

\subsection*{Data and code availability}

The cycle-accurate simulator, RTL sources, training scripts, tournament transcripts, analysis code and plotting scripts, and trained crossbar weight matrices will be deposited in a public repository (Zenodo) upon publication, with the persistent digital object identifier (DOI) provided in the published version. The repository will include all code required to reproduce the figures, tables and numerical claims in this manuscript. Reviewer access via an anonymized private link is available on request to the corresponding author.


\backmatter

\bmhead{Supplementary information}
The Supplementary Information accompanies this manuscript and contains: (1) MCTS algorithm details; (2) phase-to-primitive decomposition methodology in full; (3) RTL design and synthesis methodology; (4) per-component circuit details for CAM, GLU, RRAM crossbar, in-memory backpropagation, and FSM controller; (5) crossbar noise-robustness analysis; (6) baseline implementations and methodology; (7) ablation studies; (8) hardware comparison with prior MCTS accelerators; (9) per-application encodings, training protocols and Search Guidance Gain definition; (10) edge-deployment scenarios; supplementary tables and figures.

\bmhead{Acknowledgments}
We thank colleagues at the Duke Center for Computational Evolutionary Intelligence (CEI) and at Hewlett Packard Labs (HPL) for technical discussions throughout this work. Crossbar device parameters were drawn from previously fabricated $64{\times}64$ arrays at HPL; we thank the HPL fabrication and characterization teams for making these data available. Computational resources were provided by Hewlett Packard Enterprise (HPE) compute infrastructures.

\section*{Declarations}

\textbf{Funding.} This work was supported by Hewlett Packard Labs and by the U.S.\ Department of Energy under Award Nos.\ DE-SC0026254 and DE-SC0026382, the National Science Foundation under Award Nos.\ 2328805 and 2328712, and the Air Force Office of Scientific Research under Award No.\ FA9550-24-1-0322. The funders had no role in study design, data collection and analysis, decision to publish, or preparation of the manuscript.

%
%
\textbf{Competing interests.} T.M.-O.\ and A.N.\ are named inventors on a pending U.S.\ patent application filed by Hewlett Packard Enterprise Development LP relating to the in-memory MCTS acceleration architecture described in this manuscript. A.G., G.P., J.I.\ and A.N.\ are employees of Hewlett Packard Labs. The remaining authors declare no competing interests.

\textbf{Author contributions.} 
T.M.-O.\ conceived the phase-to-primitive decomposition methodology, designed the IMC-MCTS architecture, performed RTL synthesis and cycle-accurate simulation, conducted the Go tournament and cross-domain evaluation, and led the manuscript drafting. B.F.M.\ contributed to architecture-level design and manuscript drafting. Y.H.\ provided guidance on architecture-level design decisions and contributed to manuscript drafting and revision. A.G.\ provided GPU benchmarking and profiling. G.P. contributed to the manuscript drafting and discussions on rollout unit at HPE Labs. H.L.\ provided guidance on memory-system trade-offs. Y.C.\ co-supervised the work overall. J.I.\ co-supervised the HPE-side contributions. A.N.\ co-supervised the HPE-side contributions and the conception of the work with the architecture design and coordinated the overall effort. All authors reviewed and approved the manuscript.

\textbf{Data availability.} See Methods.

\textbf{Code availability.} See Methods.


\begin{appendices}

\section*{Extended Data}

\setcounter{figure}{0}
\renewcommand{\figurename}{Extended Data Fig.}
\setcounter{table}{0}
\renewcommand{\tablename}{Extended Data Table}


\begin{figure}[!htbp]
\centering
\includegraphics[width=0.85\linewidth,height=6cm,keepaspectratio]{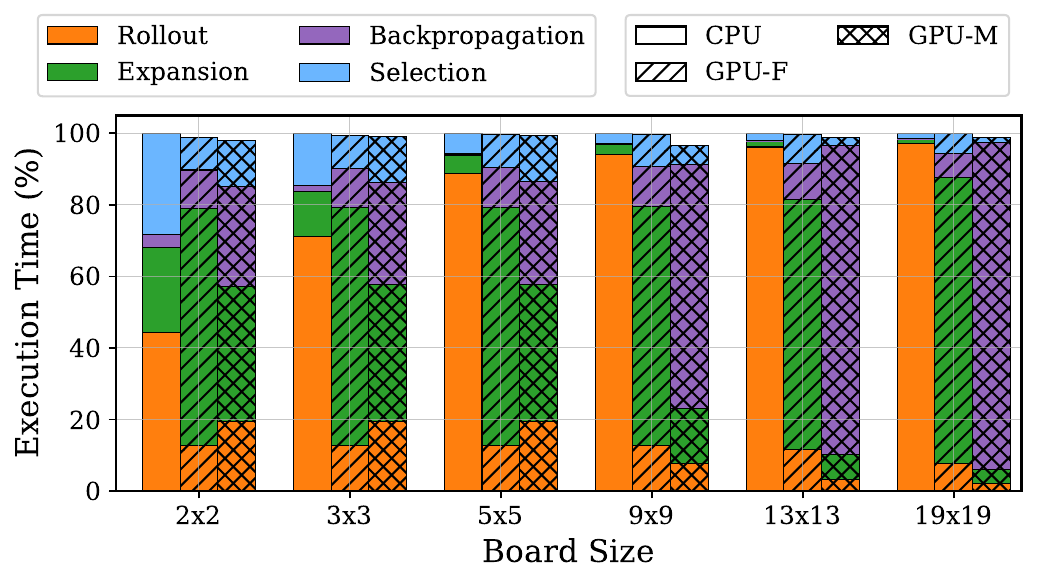}
\caption{\textbf{Phase-wise execution-time profiling.} Per-phase fraction of total MCTS execution time on a central processing unit (CPU), graphics processing unit (GPU)-Fair (1 tree, 1 playout) and GPU-Max (8 trees, 128 playouts) across six Go board sizes ($2{\times}2$ to $19{\times}19$).}
\label{eda:bottlenecks}
\end{figure}


\begin{table*}[!htbp]
\centering
\caption{\textbf{MCTS phase breakdown on central processing unit (CPU) vs.\ graphics processing unit (GPU) at $9{\times}9$ Go (5{,}000 iterations).} Bold cells mark the per-platform bottleneck. Measured on AMD Ryzen Threadripper PRO 5945WX and NVIDIA H100 80\,GB High Bandwidth Memory 3 (HBM3). The traditional-MCTS backpropagation row (2{,}482\,ms, 68.2\%) reflects per-iteration Compute Unified Device Architecture (CUDA) kernel-launch overhead and global-memory counter-update contention in the \texttt{mcts\_numba\_cuda} reference; neural-MCTS amortizes the same operation within larger CUDA Basic Linear Algebra Subprograms (cuBLAS) evaluator batches and so reports 0.14\,ms on identical hardware. The Other/overhead row collects scheduler, launch, measurement and rounding overhead not attributable to one MCTS phase.}
\label{edt:phase_breakdown}
\footnotesize
\resizebox{\textwidth}{!}{%
\begin{tabular}{@{}lccccc@{}}
\toprule
\textbf{Phase} & \multicolumn{2}{c}{\textbf{Traditional MCTS}} & \multicolumn{2}{c}{\textbf{Neural MCTS}} & \textbf{Bottleneck} \\
\cmidrule(lr){2-3} \cmidrule(lr){4-5}
               & \textbf{CPU (AMD)} & \textbf{GPU-M (H100)} & \textbf{CPU (AMD)} & \textbf{GPU-M (H100)} & \\
\midrule
Selection      & 5.88\,ms (5.3\%)     & 188.5\,ms (5.2\%)         & 7.43\,ms (11.6\%)    & 4.04\,ms (1.7\%)      & Hash lookup \\
Expansion      & 8.68\,ms (7.8\%)     & 556.8\,ms (15.3\%)        & 5.63\,ms (8.8\%)     & 7.42\,ms (3.1\%)      & Thread divergence \\
Rollout        & \textbf{95.5\,ms (86.3\%)}    & 281.6\,ms (7.7\%)         & \textbf{50.6\,ms (78.8\%)}    & \textbf{227.8\,ms (95.0\%)}    & Compute-bound \\
Backprop       & 0.11\,ms (0.1\%)     & \textbf{2{,}482.2\,ms (68.2\%)}       & 0.12\,ms (0.2\%)     & 0.14\,ms (0.1\%)      & Memory bandwidth \\
Other/overhead & 0.43\,ms (0.4\%)     & 129.9\,ms (3.6\%)         & 0.52\,ms (0.8\%)     & 0.40\,ms (0.2\%)      & Scheduler/launch \\
\midrule
\textbf{Total} & \textbf{110.6\,ms}   & \textbf{3{,}639.0\,ms}       & \textbf{64.3\,ms}    & \textbf{239.8\,ms}    & --- \\
\bottomrule
\end{tabular}%
}
\end{table*}

\begin{table}[!htbp]
\centering
\caption{\textbf{Energy per 5{,}000-iteration search across board sizes at medium play strength.} Neural-network (NN) evaluation in all configurations. CPU-NN (CPU with NN evaluator) and GPU-NN (GPU with NN evaluator) values are mean $\pm$ s.d.\ across 5 trials; IMC-NN (IMC with NN evaluator) values are from cycle-accurate deterministic simulation (single run, no variance) and therefore reported without error bars. Comparisons start at $3{\times}3$; the $2{\times}2$ GPU-NN measurement was dominated by H100 warm-up and idle-power floor and is omitted.}
\label{edt:scaling}
\footnotesize
\begin{tabular}{lcccc}
\toprule
\textbf{Board} & \textbf{CPU-NN} & \textbf{GPU-NN} & \textbf{IMC-NN} & \textbf{IMC vs.} \\
\textbf{Size}  & \textbf{(mJ)}   & \textbf{(mJ)}   & \textbf{(mJ)}   & \textbf{GPU-NN} \\
\midrule
$3{\times}3$    & 86.1 $\pm$ 78.6         & 16{,}523 $\pm$ 155      & 0.14        & 118{,}021$\times$ \\
$5{\times}5$    & 390.3 $\pm$ 84.4        & 33{,}171 $\pm$ 296      & 1.40        & 23{,}693$\times$ \\
$9{\times}9$    & 8{,}471 $\pm$ 471       & 167{,}878 $\pm$ 2{,}535 & 124.18      & \textbf{1{,}351$\times$} \\
$13{\times}13$  & 51{,}349 $\pm$ 3{,}975  & 260{,}477 $\pm$ 1{,}571 & 1{,}144.66  & 228$\times$ \\
$19{\times}19$  & 157{,}133 $\pm$ 7{,}980 & 370{,}789 $\pm$ 1{,}692 & 12{,}865.28 & 29$\times$ \\
\bottomrule
\end{tabular}
\end{table}

\begin{table*}[!htbp]
\centering
\caption{\textbf{Cross-domain generalizability across 8 applications spanning 4 AI domains.} Per-application encoding, crossbar dimensions, evaluator accuracy (MCTS+neural network (NN)), area, energy per move and Search Guidance Gain (SGG; Methods). Hardware modeled at 22\,nm. Per-application customization is a single-pass digital synthesis: the parameterized SystemVerilog design (Methods, \emph{IMC-MCTS architecture and synthesis}) is re-synthesized with the application's board-size and encoding parameters, which sizes the CAM, GLU, SRAM and crossbar arrays accordingly; no RTL logic is rewritten per application. At runtime, only the crossbar weights are reprogrammed.}
\label{edt:generalisation}
\footnotesize
\resizebox{\textwidth}{!}{%
\begin{tabular}{ll l c c c cc c}
\toprule
\textbf{Domain} & \textbf{Application} & \textbf{Encoding} & \textbf{Board} & \textbf{Crossbar} & \textbf{MCTS+NN} & \textbf{Area} & \textbf{Energy} & \textbf{SGG} \\
& & & \textbf{Size} & \textbf{(In$\to$H$\to$Out)} & \textbf{(\%)} & \textbf{(mm$^2$)} & \textbf{($\mu$J)} & \\
\midrule
\multicolumn{9}{c}{\textit{Board games (2-player; head-to-head win rate, $N=10$, sides swapped)}} \\
\midrule
Strategy & Connect Four$^\dagger$    & \{empty,R,Y\}                    & 8      & 129$\to$96$\to$3   & 80          & 1.30 & 2.6  & +0.37 \\
Strategy & Othello$^\dagger$         & \{empty,B,W\}                    & 8      & 129$\to$96$\to$3   & 80          & 1.30 & 2.6  & +0.30 \\
Strategy & Hex$^\ddagger$             & stone + edge dist.               & 11     & 485$\to$96$\to$3   & 80          & 1.87 & 9.3  & +0.22 \\
Strategy & Go                       & \{empty,B,W\}                    & 9      & 162$\to$96$\to$3   & 60          & 1.84 & 4.2  & +0.09 \\
\midrule
\multicolumn{9}{c}{\textit{Single-player (avg.\ score; SGG uses score lift over random-MCTS baseline)}} \\
\midrule
Navigation & FrozenLake$^\dagger$    & \{safe,hole,goal\}               & 8      & 129$\to$96$\to$3   & 100         & 1.30 & 2.6  & +0.15 \\
Navigation & MiniGrid               & \{empty,wall,goal\}              & 8      & 129$\to$96$\to$3   & 100         & 1.30 & 2.6  & +0.56 \\
Science    & HP Protein Fold$^\ddagger$& \{empty,H,P\} + seq.\ ctx.       & 13     & 344$\to$128$\to$3  & 100         & 2.79 & 17.8 & +0.20 \\
Puzzle     & Minesweeper$^\ddagger$   & cell state + adj.\ count         & 9      & 244$\to$96$\to$3   & 93          & 1.84 & 4.2  & +0.11 \\
\bottomrule
\end{tabular}%
}
\\[0.3em]
{\scriptsize $^{\dagger}$ Empty-position detection generates a superset of legal moves; the network suppresses illegal ones. $\quad$ $^{\ddagger}$ Encoding extends beyond three-state representation (see column).}
\end{table*}

\begin{table*}[!htbp]
\centering
\caption{\textbf{Representative prior MCTS hardware accelerators.} Five anchor entries from a 12-work survey across field-programmable gate array (FPGA), graphics processing unit (GPU), neuromorphic and multi-core central processing unit (CPU) platforms (full survey in Supplementary Table~S6); application-specific integrated circuit (ASIC) denotes a custom chip platform. Energy-efficiency entries report the work's stated figure against its own baseline; ``---'' denotes no energy figure reported. No prior work uses in-memory computing for any MCTS phase; 11 of 12 surveyed works retain random-simulation rollouts (only Thomas 2022 uses a neural evaluator); only Ho et al.\ 2024 reports an energy-efficiency figure.}
\label{edt:prior_work}
\footnotesize
\begin{tabular}{lllc}
\toprule
\textbf{Work} & \textbf{Platform} & \textbf{Rollout} & \textbf{Energy efficiency} \\
\midrule
Meng et al.\ 2023\cite{meng2023framework}     & CPU--FPGA           & Random sim.            & --- \\
Buzer \& Cazenave 2023\cite{buzer2023mcts}    & GPU                 & Random sim.            & --- \\
Thomas 2022\cite{thomas2022julia}             & GPU                 & NN evaluator           & --- \\
Ho et al.\ 2024\cite{ho2024neuromorphic}      & Neuromorphic        & Random sim.            & 5.6--10.4$\times$ \\
Steinmetz \& Gini 2020\cite{steinmetz2020tog} & Multi-core CPU      & Random sim.            & --- \\
\midrule
\textbf{This work}                            & \textbf{ASIC + IMC} & \textbf{NN evaluator}  & \textbf{\shortstack{96$\times$ over CPU \\ 65$\times$--2{,}059$\times$ over H100}} \\
\bottomrule
\end{tabular}
\end{table*}

\end{appendices}


\newpage
\bibliography{sn-bibliography}


\begin{thebibliography}{42}
\ifx \bisbn   \undefined \def \bisbn  #1{ISBN #1}\fi
\ifx \binits  \undefined \def \binits#1{#1}\fi
\ifx \bauthor  \undefined \def \bauthor#1{#1}\fi
\ifx \batitle  \undefined \def \batitle#1{#1}\fi
\ifx \bjtitle  \undefined \def \bjtitle#1{#1}\fi
\ifx \bvolume  \undefined \def \bvolume#1{\textbf{#1}}\fi
\ifx \byear  \undefined \def \byear#1{#1}\fi
\ifx \bissue  \undefined \def \bissue#1{#1}\fi
\ifx \bfpage  \undefined \def \bfpage#1{#1}\fi
\ifx \blpage  \undefined \def \blpage #1{#1}\fi
\ifx \burl  \undefined \def \burl#1{\textsf{#1}}\fi
\ifx \doiurl  \undefined \def \doiurl#1{\url{https://doi.org/#1}}\fi
\ifx \betal  \undefined \def \betal{\textit{et al.}}\fi
\ifx \binstitute  \undefined \def \binstitute#1{#1}\fi
\ifx \binstitutionaled  \undefined \def \binstitutionaled#1{#1}\fi
\ifx \bctitle  \undefined \def \bctitle#1{#1}\fi
\ifx \beditor  \undefined \def \beditor#1{#1}\fi
\ifx \bpublisher  \undefined \def \bpublisher#1{#1}\fi
\ifx \bbtitle  \undefined \def \bbtitle#1{#1}\fi
\ifx \bedition  \undefined \def \bedition#1{#1}\fi
\ifx \bseriesno  \undefined \def \bseriesno#1{#1}\fi
\ifx \blocation  \undefined \def \blocation#1{#1}\fi
\ifx \bsertitle  \undefined \def \bsertitle#1{#1}\fi
\ifx \bsnm \undefined \def \bsnm#1{#1}\fi
\ifx \bsuffix \undefined \def \bsuffix#1{#1}\fi
\ifx \bparticle \undefined \def \bparticle#1{#1}\fi
\ifx \barticle \undefined \def \barticle#1{#1}\fi
\bibcommenthead
\ifx \bconfdate \undefined \def \bconfdate #1{#1}\fi
\ifx \botherref \undefined \def \botherref #1{#1}\fi
\ifx \url \undefined \def \url#1{\textsf{#1}}\fi
\ifx \bchapter \undefined \def \bchapter#1{#1}\fi
\ifx \bbook \undefined \def \bbook#1{#1}\fi
\ifx \bcomment \undefined \def \bcomment#1{#1}\fi
\ifx \oauthor \undefined \def \oauthor#1{#1}\fi
\ifx \citeauthoryear \undefined \def \citeauthoryear#1{#1}\fi
\ifx \endbibitem  \undefined \def \endbibitem {}\fi
\ifx \bconflocation  \undefined \def \bconflocation#1{#1}\fi
\ifx \arxivurl  \undefined \def \arxivurl#1{\textsf{#1}}\fi
\csname PreBibitemsHook\endcsname

\bibitem[\protect\citeauthoryear{Coulom}{2006}]{coulom2006efficient}
\begin{bchapter}
\bauthor{\bsnm{Coulom}, \binits{R.}}:
\bctitle{Efficient selectivity and backup operators in {M}onte-{C}arlo tree
  search}.
In: \bbtitle{International Conference on Computers and Games},
pp. \bfpage{72}--\blpage{83}.
\bpublisher{Springer},
\blocation{Berlin, Heidelberg}
(\byear{2006})
\end{bchapter}
\endbibitem

\bibitem[\protect\citeauthoryear{Kocsis and Szepesv{\'a}ri}{2006}]{uct}
\begin{bchapter}
\bauthor{\bsnm{Kocsis}, \binits{L.}},
\bauthor{\bsnm{Szepesv{\'a}ri}, \binits{C.}}:
\bctitle{Bandit based {M}onte-{C}arlo planning}.
In: \bbtitle{European Conference on Machine Learning (ECML)},
pp. \bfpage{282}--\blpage{293}.
\bpublisher{Springer},
\blocation{Berlin, Heidelberg}
(\byear{2006})
\end{bchapter}
\endbibitem

\bibitem[\protect\citeauthoryear{Silver et~al.}{2016}]{alphago}
\begin{barticle}
\bauthor{\bsnm{Silver}, \binits{D.}},
\bauthor{\bsnm{Huang}, \binits{A.}},
\bauthor{\bsnm{Maddison}, \binits{C.J.}},
\bauthor{\bsnm{Guez}, \binits{A.}},
\bauthor{\bsnm{Sifre}, \binits{L.}},
\bauthor{\bsnm{Driessche}, \binits{G.}},
\bauthor{\bsnm{Schrittwieser}, \binits{J.}},
\bauthor{\bsnm{Antonoglou}, \binits{I.}},
\bauthor{\bsnm{Panneershelvam}, \binits{V.}},
\bauthor{\bsnm{Lanctot}, \binits{M.}},
\bauthor{\bsnm{Dieleman}, \binits{S.}},
\bauthor{\bsnm{Grewe}, \binits{D.}},
\bauthor{\bsnm{Nham}, \binits{J.}},
\bauthor{\bsnm{Kalchbrenner}, \binits{N.}},
\bauthor{\bsnm{Sutskever}, \binits{I.}},
\bauthor{\bsnm{Lillicrap}, \binits{T.}},
\bauthor{\bsnm{Leach}, \binits{M.}},
\bauthor{\bsnm{Kavukcuoglu}, \binits{K.}},
\bauthor{\bsnm{Graepel}, \binits{T.}},
\bauthor{\bsnm{Hassabis}, \binits{D.}}:
\batitle{Mastering the game of {Go} with deep neural networks and tree search}.
\bjtitle{Nature}
\bvolume{529}(\bissue{7587}),
\bfpage{484}--\blpage{489}
(\byear{2016})
\doiurl{10.1038/nature16961}
\end{barticle}
\endbibitem

\bibitem[\protect\citeauthoryear{Silver et~al.}{2018}]{silver2017alphazero}
\begin{barticle}
\bauthor{\bsnm{Silver}, \binits{D.}},
\bauthor{\bsnm{Hubert}, \binits{T.}},
\bauthor{\bsnm{Schrittwieser}, \binits{J.}},
\bauthor{\bsnm{Antonoglou}, \binits{I.}},
\bauthor{\bsnm{Lai}, \binits{M.}},
\bauthor{\bsnm{Guez}, \binits{A.}},
\bauthor{\bsnm{Lanctot}, \binits{M.}},
\bauthor{\bsnm{Sifre}, \binits{L.}},
\bauthor{\bsnm{Kumaran}, \binits{D.}},
\bauthor{\bsnm{Graepel}, \binits{T.}},
\bauthor{\bsnm{Lillicrap}, \binits{T.P.}},
\bauthor{\bsnm{Simonyan}, \binits{K.}},
\bauthor{\bsnm{Hassabis}, \binits{D.}}:
\batitle{A general reinforcement learning algorithm that masters chess, shogi,
  and {Go} through self-play}.
\bjtitle{Science}
\bvolume{362}(\bissue{6419}),
\bfpage{1140}--\blpage{1144}
(\byear{2018})
\doiurl{10.1126/science.aar6404}
\end{barticle}
\endbibitem

\bibitem[\protect\citeauthoryear{Dam et~al.}{2022}]{dam2022robot}
\begin{barticle}
\bauthor{\bsnm{Dam}, \binits{T.}},
\bauthor{\bsnm{Chalvatzaki}, \binits{G.}},
\bauthor{\bsnm{Peters}, \binits{J.}},
\bauthor{\bsnm{Pajarinen}, \binits{J.}}:
\batitle{{M}onte-{C}arlo robot path planning}.
\bjtitle{IEEE Robotics and Automation Letters}
\bvolume{7}(\bissue{4}),
\bfpage{11213}--\blpage{11220}
(\byear{2022})
\doiurl{10.1109/LRA.2022.3199674}
\end{barticle}
\endbibitem

\bibitem[\protect\citeauthoryear{Deng et~al.}{2022}]{deng2022protein}
\begin{barticle}
\bauthor{\bsnm{Deng}, \binits{H.}},
\bauthor{\bsnm{Yuan}, \binits{X.}},
\bauthor{\bsnm{Tian}, \binits{Y.}},
\bauthor{\bsnm{Hu}, \binits{J.}}:
\batitle{Neural-augmented two-stage {M}onte {C}arlo tree search with
  over-sampling for protein folding in {HP} model}.
\bjtitle{IEEJ Transactions on Electrical and Electronic Engineering}
\bvolume{17}(\bissue{5}),
\bfpage{685}--\blpage{694}
(\byear{2022})
\doiurl{10.1002/tee.23556}
\end{barticle}
\endbibitem

\bibitem[\protect\citeauthoryear{Buzer and Cazenave}{2023}]{buzer2023mcts}
\begin{bchapter}
\bauthor{\bsnm{Buzer}, \binits{L.}},
\bauthor{\bsnm{Cazenave}, \binits{T.}}:
\bctitle{{GPU} for {M}onte {C}arlo search}.
In: \bbtitle{International Conference on Learning and Intelligent Optimization
  (LION)}.
\bsertitle{Lecture Notes in Computer Science},
vol. \bseriesno{14286},
pp. \bfpage{179}--\blpage{193}.
\bpublisher{Springer},
\blocation{Cham}
(\byear{2023}).
\doiurl{10.1007/978-3-031-44505-7_13}
\end{bchapter}
\endbibitem

\bibitem[\protect\citeauthoryear{{Intel Corporation}}{2023}]{intel_xeon_tdp}
\begin{botherref}
\oauthor{\bsnm{{Intel Corporation}}}:
{Intel Xeon Platinum 8462Y+} Processor Specifications.
Product brief
(2023)
\end{botherref}
\endbibitem

\bibitem[\protect\citeauthoryear{{NVIDIA Corporation}}{2022}]{nvidia_gpu_power}
\begin{botherref}
\oauthor{\bsnm{{NVIDIA Corporation}}}:
{NVIDIA H100} Tensor Core GPU Architecture.
Whitepaper
(2022)
\end{botherref}
\endbibitem

\bibitem[\protect\citeauthoryear{Chaslot et~al.}{2008}]{chaslot2008parallel}
\begin{bchapter}
\bauthor{\bsnm{Chaslot}, \binits{G.M.J.-B.}},
\bauthor{\bsnm{Winands}, \binits{M.H.M.}},
\bauthor{\bsnm{Herik}, \binits{H.J.}}:
\bctitle{Parallel {M}onte-{C}arlo tree search}.
In: \bbtitle{International Conference on Computers and Games (CG 2008)},
pp. \bfpage{60}--\blpage{71}.
\bpublisher{Springer},
\blocation{Berlin, Heidelberg}
(\byear{2008})
\end{bchapter}
\endbibitem

\bibitem[\protect\citeauthoryear{Rocki and Suda}{2011}]{rocki2011parallel}
\begin{bchapter}
\bauthor{\bsnm{Rocki}, \binits{K.}},
\bauthor{\bsnm{Suda}, \binits{R.}}:
\bctitle{Large-scale parallel {M}onte-{C}arlo tree search on {GPU}}.
In: \bbtitle{2011 IEEE International Symposium on Parallel and Distributed
  Processing Workshops and PhD Forum},
pp. \bfpage{2034}--\blpage{2037}.
\bpublisher{IEEE},
\blocation{Piscataway, NJ}
(\byear{2011})
\end{bchapter}
\endbibitem

\bibitem[\protect\citeauthoryear{Ielmini and Wong}{2018}]{ielmini2018memory}
\begin{barticle}
\bauthor{\bsnm{Ielmini}, \binits{D.}},
\bauthor{\bsnm{Wong}, \binits{H.-S.P.}}:
\batitle{In-memory computing with resistive switching devices}.
\bjtitle{Nature Electronics}
\bvolume{1}(\bissue{6}),
\bfpage{333}--\blpage{343}
(\byear{2018})
\doiurl{10.1038/s41928-018-0092-2}
\end{barticle}
\endbibitem

\bibitem[\protect\citeauthoryear{Wan et~al.}{2022}]{wan2022nature}
\begin{barticle}
\bauthor{\bsnm{Wan}, \binits{W.}},
\bauthor{\bsnm{Kubendran}, \binits{R.}},
\bauthor{\bsnm{Schaefer}, \binits{C.}},
\bauthor{\bsnm{Eryilmaz}, \binits{S.B.}},
\bauthor{\bsnm{Zhang}, \binits{W.}},
\bauthor{\bsnm{Wu}, \binits{D.}},
\bauthor{\bsnm{Deiss}, \binits{S.}},
\bauthor{\bsnm{Raina}, \binits{P.}},
\bauthor{\bsnm{Qian}, \binits{H.}},
\bauthor{\bsnm{Gao}, \binits{B.}},
\bauthor{\bsnm{Joshi}, \binits{S.}},
\bauthor{\bsnm{Wu}, \binits{H.}},
\bauthor{\bsnm{Wong}, \binits{H.-S.P.}},
\bauthor{\bsnm{Cauwenberghs}, \binits{G.}}:
\batitle{A compute-in-memory chip based on resistive random-access memory}.
\bjtitle{Nature}
\bvolume{608},
\bfpage{504}--\blpage{512}
(\byear{2022})
\doiurl{10.1038/s41586-022-04992-8}
\end{barticle}
\endbibitem

\bibitem[\protect\citeauthoryear{B{\"u}chel et~al.}{2024}]{sebastian2024kernel}
\begin{barticle}
\bauthor{\bsnm{B{\"u}chel}, \binits{J.}},
\bauthor{\bsnm{Camposampiero}, \binits{G.}},
\bauthor{\bsnm{Vasilopoulos}, \binits{A.}},
\bauthor{\bsnm{Lammie}, \binits{C.}},
\bauthor{\bsnm{Le~Gallo}, \binits{M.}},
\bauthor{\bsnm{Rahimi}, \binits{A.}},
\bauthor{\bsnm{Sebastian}, \binits{A.}}:
\batitle{Kernel approximation using analogue in-memory computing}.
\bjtitle{Nature Machine Intelligence}
(\byear{2024})
\doiurl{10.1038/s42256-024-00943-2}
\end{barticle}
\endbibitem

\bibitem[\protect\citeauthoryear{Li et~al.}{2019}]{li2019lstm}
\begin{barticle}
\bauthor{\bsnm{Li}, \binits{C.}},
\bauthor{\bsnm{Wang}, \binits{Z.}},
\bauthor{\bsnm{Rao}, \binits{M.}},
\bauthor{\bsnm{Belkin}, \binits{D.}},
\bauthor{\bsnm{Song}, \binits{W.}},
\bauthor{\bsnm{Jiang}, \binits{H.}},
\bauthor{\bsnm{Yan}, \binits{P.}},
\bauthor{\bsnm{Li}, \binits{Y.}},
\bauthor{\bsnm{Lin}, \binits{P.}},
\bauthor{\bsnm{Hu}, \binits{M.}},
\bauthor{\bsnm{Ge}, \binits{N.}},
\bauthor{\bsnm{Strachan}, \binits{J.P.}},
\bauthor{\bsnm{Barnell}, \binits{M.}},
\bauthor{\bsnm{Wu}, \binits{Q.}},
\bauthor{\bsnm{Williams}, \binits{R.S.}},
\bauthor{\bsnm{Yang}, \binits{J.J.}},
\bauthor{\bsnm{Xia}, \binits{Q.}}:
\batitle{Long short-term memory networks in memristor crossbar arrays}.
\bjtitle{Nature Machine Intelligence}
\bvolume{1},
\bfpage{49}--\blpage{57}
(\byear{2019})
\doiurl{10.1038/s42256-018-0001-4}
\end{barticle}
\endbibitem

\bibitem[\protect\citeauthoryear{Portner et~al.}{2025}]{portner2025actor}
\begin{barticle}
\bauthor{\bsnm{Portner}, \binits{K.}},
\bauthor{\bsnm{Zellweger}, \binits{T.}},
\bauthor{\bsnm{Martinelli}, \binits{F.}},
\bauthor{\bsnm{B{\'e}gon-Lours}, \binits{L.}},
\bauthor{\bsnm{Bragaglia}, \binits{V.}},
\bauthor{\bsnm{Weilenmann}, \binits{C.}},
\bauthor{\bsnm{Jubin}, \binits{D.}},
\bauthor{\bsnm{Falcone}, \binits{D.}},
\bauthor{\bsnm{Hermann}, \binits{F.}},
\bauthor{\bsnm{Hrynkevych}, \binits{O.}},
\bauthor{\bsnm{Stecconi}, \binits{T.}},
\bauthor{\bsnm{La~Porta}, \binits{A.}},
\bauthor{\bsnm{Drechsler}, \binits{U.}},
\bauthor{\bsnm{Olziersky}, \binits{A.}},
\bauthor{\bsnm{Offrein}, \binits{B.J.}},
\bauthor{\bsnm{Gerstner}, \binits{W.}},
\bauthor{\bsnm{Luisier}, \binits{M.}},
\bauthor{\bsnm{Emboras}, \binits{A.}}:
\batitle{Actor--critic networks with analogue memristors mimicking reward-based
  learning}.
\bjtitle{Nature Machine Intelligence}
(\byear{2025})
\doiurl{10.1038/s42256-025-01149-w}
\end{barticle}
\endbibitem

\bibitem[\protect\citeauthoryear{Marr}{1971}]{marr1971archicortex}
\begin{barticle}
\bauthor{\bsnm{Marr}, \binits{D.}}:
\batitle{Simple memory: A theory for archicortex}.
\bjtitle{Philosophical Transactions of the Royal Society of London. Series B,
  Biological Sciences}
\bvolume{262}(\bissue{841}),
\bfpage{23}--\blpage{81}
(\byear{1971})
\doiurl{10.1098/rstb.1971.0078}
\end{barticle}
\endbibitem

\bibitem[\protect\citeauthoryear{Treves and Rolls}{1994}]{treves1994ca3}
\begin{barticle}
\bauthor{\bsnm{Treves}, \binits{A.}},
\bauthor{\bsnm{Rolls}, \binits{E.T.}}:
\batitle{Computational analysis of the role of the hippocampus in memory}.
\bjtitle{Hippocampus}
\bvolume{4}(\bissue{3}),
\bfpage{374}--\blpage{391}
(\byear{1994})
\doiurl{10.1002/hipo.450040319}
\end{barticle}
\endbibitem

\bibitem[\protect\citeauthoryear{Mead}{1990}]{mead1990neuromorphic}
\begin{barticle}
\bauthor{\bsnm{Mead}, \binits{C.}}:
\batitle{Neuromorphic electronic systems}.
\bjtitle{Proceedings of the IEEE}
\bvolume{78}(\bissue{10}),
\bfpage{1629}--\blpage{1636}
(\byear{1990})
\doiurl{10.1109/5.58356}
\end{barticle}
\endbibitem

\bibitem[\protect\citeauthoryear{Sebastian et~al.}{2020}]{sebastian2020imc}
\begin{barticle}
\bauthor{\bsnm{Sebastian}, \binits{A.}},
\bauthor{\bsnm{Le~Gallo}, \binits{M.}},
\bauthor{\bsnm{Khaddam-Aljameh}, \binits{R.}},
\bauthor{\bsnm{Eleftheriou}, \binits{E.}}:
\batitle{Memory devices and applications for in-memory computing}.
\bjtitle{Nature Nanotechnology}
\bvolume{15},
\bfpage{529}--\blpage{544}
(\byear{2020})
\doiurl{10.1038/s41565-020-0655-z}
\end{barticle}
\endbibitem

\bibitem[\protect\citeauthoryear{Pagiamtzis and
  Sheikholeslami}{2006}]{pagiamtzis2006content}
\begin{barticle}
\bauthor{\bsnm{Pagiamtzis}, \binits{K.}},
\bauthor{\bsnm{Sheikholeslami}, \binits{A.}}:
\batitle{Content-addressable memory ({CAM}) circuits and architectures: A
  tutorial and survey}.
\bjtitle{IEEE Journal of Solid-State Circuits}
\bvolume{41}(\bissue{3}),
\bfpage{712}--\blpage{727}
(\byear{2006})
\end{barticle}
\endbibitem

\bibitem[\protect\citeauthoryear{Hu et~al.}{2018}]{hu2018memristor}
\begin{barticle}
\bauthor{\bsnm{Hu}, \binits{M.}},
\bauthor{\bsnm{Graves}, \binits{C.E.}},
\bauthor{\bsnm{Li}, \binits{C.}},
\bauthor{\bsnm{Li}, \binits{Y.}},
\bauthor{\bsnm{Ge}, \binits{N.}},
\bauthor{\bsnm{Montgomery}, \binits{E.}},
\bauthor{\bsnm{D{\'a}vila}, \binits{N.}},
\bauthor{\bsnm{Jiang}, \binits{H.}},
\bauthor{\bsnm{Williams}, \binits{R.S.}},
\bauthor{\bsnm{Yang}, \binits{J.J.}},
\bauthor{\bsnm{Xia}, \binits{Q.}},
\bauthor{\bsnm{Strachan}, \binits{J.P.}}:
\batitle{Memristor-based analog computation and neural network classification
  with a dot product engine}.
\bjtitle{Advanced Materials}
\bvolume{30}(\bissue{9}),
\bfpage{1705914}
(\byear{2018})
\end{barticle}
\endbibitem

\bibitem[\protect\citeauthoryear{Li et~al.}{2020}]{li_cmos-integrated_2020}
\begin{bchapter}
\bauthor{\bsnm{Li}, \binits{C.}},
\bauthor{\bsnm{Ignowski}, \binits{J.}},
\bauthor{\bsnm{Sheng}, \binits{X.}},
\bauthor{\bsnm{Wessel}, \binits{R.}},
\bauthor{\bsnm{Jaffe}, \binits{B.}},
\bauthor{\bsnm{Ingemi}, \binits{J.}},
\bauthor{\bsnm{Graves}, \binits{C.}},
\bauthor{\bsnm{Strachan}, \binits{J.P.}}:
\bctitle{{CMOS}-integrated nanoscale memristive crossbars for {CNN} and
  optimization acceleration}.
In: \bbtitle{2020 IEEE International Memory Workshop (IMW)},
pp. \bfpage{1}--\blpage{4}.
\bpublisher{IEEE},
\blocation{Piscataway, NJ}
(\byear{2020}).
\doiurl{10.1109/IMW48823.2020.9108112}
\end{bchapter}
\endbibitem

\bibitem[\protect\citeauthoryear{Tromp}{2016}]{tromp2016number}
\begin{bchapter}
\bauthor{\bsnm{Tromp}, \binits{J.}}:
\bctitle{The number of legal {Go} positions}.
In: \bbtitle{Computers and Games (CG 2016)}.
\bsertitle{Lecture Notes in Computer Science},
vol. \bseriesno{10068},
pp. \bfpage{183}--\blpage{190}.
\bpublisher{Springer},
\blocation{Cham}
(\byear{2016}).
\doiurl{10.1007/978-3-319-50935-8_17}
\end{bchapter}
\endbibitem

\bibitem[\protect\citeauthoryear{Wu}{2019}]{wu2019katago}
\begin{botherref}
\oauthor{\bsnm{Wu}, \binits{D.J.}}:
Accelerating self-play learning in {Go}.
arXiv preprint arXiv:1902.10565
(2019)
\end{botherref}
\endbibitem

\bibitem[\protect\citeauthoryear{Shafiee et~al.}{2016}]{shafiee2016isaac}
\begin{bchapter}
\bauthor{\bsnm{Shafiee}, \binits{A.}},
\bauthor{\bsnm{Nag}, \binits{A.}},
\bauthor{\bsnm{Muralimanohar}, \binits{N.}},
\bauthor{\bsnm{Balasubramonian}, \binits{R.}},
\bauthor{\bsnm{Strachan}, \binits{J.P.}},
\bauthor{\bsnm{Hu}, \binits{M.}},
\bauthor{\bsnm{Williams}, \binits{R.S.}},
\bauthor{\bsnm{Srikumar}, \binits{V.}}:
\bctitle{{ISAAC}: A convolutional neural network accelerator with in-situ
  analog arithmetic in crossbars}.
In: \bbtitle{Proceedings of the 43rd International Symposium on Computer
  Architecture (ISCA)},
pp. \bfpage{14}--\blpage{26}.
\bpublisher{IEEE Press},
\blocation{Piscataway, NJ, USA}
(\byear{2016}).
\doiurl{10.1145/3007787.3001139}
\end{bchapter}
\endbibitem

\bibitem[\protect\citeauthoryear{Muralimanohar et~al.}{2009}]{cacti}
\begin{botherref}
\oauthor{\bsnm{Muralimanohar}, \binits{N.}},
\oauthor{\bsnm{Balasubramonian}, \binits{R.}},
\oauthor{\bsnm{Jouppi}, \binits{N.P.}}:
{CACTI 6.0}: A tool to model large caches.
Technical Report HPL-2009-85,
HP Labs
(2009)
\end{botherref}
\endbibitem

\bibitem[\protect\citeauthoryear{Rodrigues et~al.}{2011}]{sst}
\begin{barticle}
\bauthor{\bsnm{Rodrigues}, \binits{A.F.}},
\bauthor{\bsnm{Hemmert}, \binits{K.S.}},
\bauthor{\bsnm{Barrett}, \binits{B.W.}},
\bauthor{\bsnm{Kersey}, \binits{C.}},
\bauthor{\bsnm{Oldfield}, \binits{R.}},
\bauthor{\bsnm{Weston}, \binits{M.}},
\bauthor{\bsnm{Risen}, \binits{R.}},
\bauthor{\bsnm{Cook}, \binits{J.}},
\bauthor{\bsnm{Rosenfeld}, \binits{P.}},
\bauthor{\bsnm{Cooper-Balis}, \binits{E.}},
\bauthor{\bsnm{Jacob}, \binits{B.}}:
\batitle{The structural simulation toolkit}.
\bjtitle{SIGMETRICS Performance Evaluation Review}
\bvolume{38}(\bissue{4}),
\bfpage{37}--\blpage{42}
(\byear{2011})
\end{barticle}
\endbibitem

\bibitem[\protect\citeauthoryear{Baudis and Gailly}{2011}]{baudis2011pachi}
\begin{bchapter}
\bauthor{\bsnm{Baudis}, \binits{P.}},
\bauthor{\bsnm{Gailly}, \binits{J.-l.}}:
\bctitle{{Pachi}: State of the art open source {Go} program}.
In: \bbtitle{Advances in Computer Games (ACG)},
pp. \bfpage{24}--\blpage{38}.
\bpublisher{Springer},
\blocation{Berlin, Heidelberg}
(\byear{2011}).
\doiurl{10.1007/978-3-642-31866-5_3}
\end{bchapter}
\endbibitem

\bibitem[\protect\citeauthoryear{Baudis}{2014}]{baudis_michi}
\begin{botherref}
\oauthor{\bsnm{Baudis}, \binits{P.}}:
Michi: {Minimalistic} {Go} {MCTS} Engine.
GitHub repository.
\url{https://github.com/pasky/michi}
(2014)
\end{botherref}
\endbibitem

\bibitem[\protect\citeauthoryear{Khan et~al.}{2018}]{rapl}
\begin{barticle}
\bauthor{\bsnm{Khan}, \binits{K.N.}},
\bauthor{\bsnm{Hirki}, \binits{M.}},
\bauthor{\bsnm{Niemi}, \binits{T.}},
\bauthor{\bsnm{Nurminen}, \binits{J.K.}},
\bauthor{\bsnm{Ou}, \binits{Z.}}:
\batitle{{RAPL} in action: Experiences in using {RAPL} for power measurements}.
\bjtitle{ACM Transactions on Modeling and Performance Evaluation of Computing
  Systems}
\bvolume{3}(\bissue{2}),
\bfpage{9}
(\byear{2018})
\doiurl{10.1145/3177754}
\end{barticle}
\endbibitem

\bibitem[\protect\citeauthoryear{Lenz et~al.}{2016}]{lenz2016driving}
\begin{bchapter}
\bauthor{\bsnm{Lenz}, \binits{D.}},
\bauthor{\bsnm{Kessler}, \binits{T.}},
\bauthor{\bsnm{Knoll}, \binits{A.}}:
\bctitle{Tactical cooperative planning for autonomous highway driving using
  {M}onte-{C}arlo tree search}.
In: \bbtitle{2016 IEEE Intelligent Vehicles Symposium (IV)},
pp. \bfpage{447}--\blpage{453}.
\bpublisher{IEEE},
\blocation{Piscataway, NJ}
(\byear{2016}).
\doiurl{10.1109/IVS.2016.7535424}
\end{bchapter}
\endbibitem

\bibitem[\protect\citeauthoryear{Fujiwara et~al.}{2022}]{fujiwara2022isscc}
\begin{bchapter}
\bauthor{\bsnm{Fujiwara}, \binits{H.}},
\bauthor{\bsnm{Mori}, \binits{H.}},
\bauthor{\bsnm{Zhao}, \binits{W.-C.}},
\bauthor{\bsnm{Chuang}, \binits{M.-C.}},
\bauthor{\bsnm{Naous}, \binits{R.}},
\bauthor{\bsnm{Chuang}, \binits{C.-K.}},
\bauthor{\bsnm{Hashizume}, \binits{T.}},
\bauthor{\bsnm{Sun}, \binits{D.}},
\bauthor{\bsnm{Lee}, \binits{C.-F.}},
\bauthor{\bsnm{Akarvardar}, \binits{K.}},
\bauthor{\bsnm{Adham}, \binits{S.}},
\bauthor{\bsnm{Chou}, \binits{T.-L.}},
\bauthor{\bsnm{Sinangil}, \binits{M.E.}},
\bauthor{\bsnm{Wang}, \binits{Y.}},
\bauthor{\bsnm{Chih}, \binits{Y.-D.}},
\bauthor{\bsnm{Chen}, \binits{Y.-H.}},
\bauthor{\bsnm{Liao}, \binits{H.-J.}},
\bauthor{\bsnm{Chang}, \binits{T.-Y.J.}}:
\bctitle{A 5-nm 254-{TOPS}/{W} 221-{TOPS}/mm$^2$ fully-digital
  computing-in-memory macro supporting wide-range dynamic-voltage-frequency
  scaling and simultaneous {MAC} and write operations}.
In: \bbtitle{2022 IEEE International Solid-State Circuits Conference (ISSCC)},
vol. \bseriesno{65},
pp. \bfpage{1}--\blpage{3}.
\bpublisher{IEEE},
\blocation{Piscataway, NJ}
(\byear{2022}).
\doiurl{10.1109/ISSCC42614.2022.9731754}
\end{bchapter}
\endbibitem

\bibitem[\protect\citeauthoryear{Cai et~al.}{2020}]{cai2020optimization}
\begin{barticle}
\bauthor{\bsnm{Cai}, \binits{F.}},
\bauthor{\bsnm{Kumar}, \binits{S.}},
\bauthor{\bsnm{Van~Vaerenbergh}, \binits{T.}},
\bauthor{\bsnm{Sheng}, \binits{X.}},
\bauthor{\bsnm{Liu}, \binits{R.}},
\bauthor{\bsnm{Li}, \binits{C.}},
\bauthor{\bsnm{Liu}, \binits{Z.}},
\bauthor{\bsnm{Foltin}, \binits{M.}},
\bauthor{\bsnm{Yu}, \binits{S.}},
\bauthor{\bsnm{Xia}, \binits{Q.}},
\bauthor{\bsnm{Yang}, \binits{J.J.}},
\bauthor{\bsnm{Beausoleil}, \binits{R.}},
\bauthor{\bsnm{Lu}, \binits{W.D.}},
\bauthor{\bsnm{Strachan}, \binits{J.P.}}:
\batitle{Power-efficient combinatorial optimization using intrinsic noise in
  memristor hopfield neural networks}.
\bjtitle{Nature Electronics}
\bvolume{3}(\bissue{7}),
\bfpage{409}--\blpage{418}
(\byear{2020})
\doiurl{10.1038/s41928-020-0436-6}
\end{barticle}
\endbibitem

\bibitem[\protect\citeauthoryear{Stillmaker and Baas}{2017}]{STILLMAKER201774}
\begin{barticle}
\bauthor{\bsnm{Stillmaker}, \binits{A.}},
\bauthor{\bsnm{Baas}, \binits{B.}}:
\batitle{Scaling equations for the accurate prediction of {CMOS} device
  performance from 180\,nm to 7\,nm}.
\bjtitle{Integration}
\bvolume{58},
\bfpage{74}--\blpage{81}
(\byear{2017})
\end{barticle}
\endbibitem

\bibitem[\protect\citeauthoryear{Auth et~al.}{2012}]{auth2012cmos}
\begin{bchapter}
\bauthor{\bsnm{Auth}, \binits{C.}}, \betal:
\bctitle{A 22nm high performance and low-power {CMOS} technology featuring
  fully-depleted tri-gate transistors, self-aligned contacts and high density
  {MIM} capacitors}.
In: \bbtitle{2012 Symposium on VLSI Technology (VLSIT)},
pp. \bfpage{131}--\blpage{132}
(\byear{2012}).
\doiurl{10.1109/VLSIT.2012.6242496}
\end{bchapter}
\endbibitem

\bibitem[\protect\citeauthoryear{Sheng et~al.}{2019}]{sheng2019low}
\begin{barticle}
\bauthor{\bsnm{Sheng}, \binits{X.}},
\bauthor{\bsnm{Graves}, \binits{C.E.}},
\bauthor{\bsnm{Kumar}, \binits{S.}},
\bauthor{\bsnm{Li}, \binits{X.}},
\bauthor{\bsnm{Buchanan}, \binits{B.}},
\bauthor{\bsnm{Zheng}, \binits{L.}},
\bauthor{\bsnm{Lam}, \binits{S.}},
\bauthor{\bsnm{Li}, \binits{C.}},
\bauthor{\bsnm{Strachan}, \binits{J.P.}}:
\batitle{Low-conductance and multilevel {CMOS}-integrated nanoscale oxide
  memristors}.
\bjtitle{Advanced Electronic Materials}
\bvolume{5}(\bissue{9}),
\bfpage{1800876}
(\byear{2019})
\end{barticle}
\endbibitem

\bibitem[\protect\citeauthoryear{Kl{\k{e}}sk}{2025}]{mctsnc_klesk2025}
\begin{barticle}
\bauthor{\bsnm{Kl{\k{e}}sk}, \binits{P.}}:
\batitle{{MCTS-NC}: A thorough {GPU} parallelization of {M}onte-{C}arlo tree
  search implemented in python via numba.cuda}.
\bjtitle{SoftwareX}
\bvolume{30},
\bfpage{102139}
(\byear{2025})
\doiurl{10.1016/j.softx.2025.102139}
\end{barticle}
\endbibitem

\bibitem[\protect\citeauthoryear{Meng et~al.}{2023}]{meng2023framework}
\begin{bchapter}
\bauthor{\bsnm{Meng}, \binits{Y.}},
\bauthor{\bsnm{Kannan}, \binits{R.}},
\bauthor{\bsnm{Prasanna}, \binits{V.}}:
\bctitle{A framework for {M}onte-{C}arlo tree search on {CPU}--{FPGA}
  heterogeneous platform via on-chip dynamic tree management}.
In: \bbtitle{Proceedings of the 2023 ACM/SIGDA International Symposium on Field
  Programmable Gate Arrays (FPGA)},
pp. \bfpage{235}--\blpage{245}.
\bpublisher{ACM},
\blocation{New York, NY, USA}
(\byear{2023}).
\doiurl{10.1145/3543622.3573177}
\end{bchapter}
\endbibitem

\bibitem[\protect\citeauthoryear{Thomas}{2022}]{thomas2022julia}
\begin{botherref}
\oauthor{\bsnm{Thomas}, \binits{G.}}:
A Full-{GPU} Implementation of {MCTS} in {Julia}: The Key to {Gumbel MuZero}?
Google Summer of Code 2022 project report (AlphaZero.jl).
Available online
(2022)
\end{botherref}
\endbibitem

\bibitem[\protect\citeauthoryear{Ho et~al.}{2024}]{ho2024neuromorphic}
\begin{bchapter}
\bauthor{\bsnm{Ho}, \binits{Y.}},
\bauthor{\bsnm{Carbajal}, \binits{A.}},
\bauthor{\bsnm{Escamilla}, \binits{L.}},
\bauthor{\bsnm{Pinar}, \binits{A.}}:
\bctitle{Neuromorphic {M}onte {C}arlo tree search methods for shortest path
  interdiction}.
In: \bbtitle{2024 International Conference on Neuromorphic Systems (ICONS)},
pp. \bfpage{307}--\blpage{311}.
\bpublisher{IEEE},
\blocation{Piscataway, NJ}
(\byear{2024}).
\doiurl{10.1109/ICONS62911.2024.00053}
\end{bchapter}
\endbibitem

\bibitem[\protect\citeauthoryear{Steinmetz and Gini}{2021}]{steinmetz2020tog}
\begin{barticle}
\bauthor{\bsnm{Steinmetz}, \binits{E.}},
\bauthor{\bsnm{Gini}, \binits{M.}}:
\batitle{More trees or larger trees: Parallelizing {M}onte {C}arlo tree
  search}.
\bjtitle{IEEE Transactions on Games}
\bvolume{13}(\bissue{3}),
\bfpage{315}--\blpage{320}
(\byear{2021})
\doiurl{10.1109/TG.2020.3048331}
\end{barticle}
\endbibitem

\end{thebibliography}



\section*{Supplementary material}

# Supplementary Information for *Multi-primitive in-memory computing for Monte Carlo tree search*

Tergel Molom-Ochir[1,2*], Benjamin F. Morris III[1], Yintao He[1], Archit Gajjar[2], Giacomo Pedretti[2], Hai "Helen" Li[1], Yiran Chen[1], Jim Ignowski[2], Aishwarya Natarajan[2]

[1*]Department of Electrical and Computer Engineering, Duke University, Durham, 27708, NC, USA.
[2]Hewlett Packard Labs, Milpitas, 95035, CA, USA.

*Corresponding author(s). E-mail(s): tergel.molom-ochir@duke.edu;
Contributing authors: ben.morris@duke.edu; yintao.he@duke.edu; archit.gajjar@hpe.com; giacomo.pedretti@hpe.com; hai.li@duke.edu; yiran.chen@duke.edu; jim.ignowski@hpe.com; aishwarya.natarajan@hpe.com;

**Abstract**

This Supplementary Information accompanies the main paper and contains: (1) MCTS algorithm details and the UCB1 / PUCT formulations; (2) the phase-to-primitive decomposition methodology in full, with the three reformulations required for MCTS; (3) RTL design and synthesis methodology; (4) per-component circuit details for CAM, GLU, RRAM crossbar, in-memory back-propagation and FSM controller; (5) crossbar noise-robustness analysis; (6) baseline implementations and methodology; (7) ablation studies; (8) comparison with prior MCTS hardware accelerators; (9) per-application encodings, training protocols, and the Search Guidance Gain definition. Cross-references in the main paper of the form "Supplementary Note ***N***" point to the corresponding section here.

# Contents



# Supplementary Note 1

# MCTS algorithm details

## 1.1 The four phases

Monte Carlo tree search (MCTS) constructs decision trees through statistical sampling via four iterative phases (Algorithm 1)[1, 2]: *selection* navigates root-to-leaf via the Upper Confidence Bound (UCB1) policy (Eq. S1), balancing exploitation and exploration; *expansion* adds child states by generating all legal moves from the current position; *rollout* evaluates the position via random simulation to a terminal state or a single neural-network forward pass; and *backpropagation* updates visit counts and win statistics along the leaf-to-root path. We evaluate on Go, where the number of legal positions reaches $10^{170}$[3], making exhaustive search intractable. MCTS executes $N$ iterations (typically 5,000) per move, each exploring one path before selecting the highest-visit-count child.

**Algorithm 1** Monte Carlo Tree Search

**Require:** Initial state $s_0$, iteration budget $N$, exploration constant $C$
**Ensure:** Best move $m^*$ from root state

```
root ← CreateNode(s_0)
for i = 1 to N do
    // Phase 1: Selection
    node ← root
    while node has children and node not terminal do
        node ← arg max_{c∈node.children} UCB1(c, C)
    end while
    // Phase 2: Expansion
    if node.visits > 0 and node not terminal then
        moves ← GetLegalMoves(node.state)
        node ← AddChild(node, RandomChoice(moves))
    end if
    // Phase 3: Rollout
    state ← node.state
    while state not terminal do
        move ← RandomMove(state); state ← ApplyMove(state, move)
    end while
    result ← Evaluate(state)
    // Phase 4: Backpropagation
    while node ≠ null do
        node.visits ← node.visits + 1; node.wins ← node.wins + result
        node ← node.parent
    end while
end for
return arg max_{c∈root.children} c.visits
```

## 1.2 UCB1 and the exploration–exploitation tradeoff

The Upper Confidence Bound (UCB1) policy balances exploitation and exploration during the selection phase:

$$\mathrm{UCB1}(v_i) = \frac{w_i}{n_i} + C\sqrt{\frac{\ln N}{n_i}}, \tag{S1}$$

where $w_i$ and $n_i$ are the wins and visits accumulated for child $i$, $N$ is the parent's visit count, and $C$ is the exploration constant ($C = \sqrt{2}$ is optimal[2]). Traditional MCTS uses random rollouts (simulating games via random legal moves to terminal states); modern implementations replace random rollouts with neural-network evaluation (a single forward pass)[4, 5]. Tree search remains essential because UCB1 prevents local optima and discovers tactical sequences that a network alone may miss[5].

## 1.3 Extension to PUCT

The selection unit extends to AlphaZero's PUCT policy[5], which replaces $\ln(N)$ with $\sqrt{N}$ and adds a neural policy prior $P(s, a_i)$:

$$\mathrm{PUCT}(v_i) = \frac{w_i}{n_i} + C \cdot P(s, a_i) \cdot \frac{\sqrt{N}}{1 + n_i}. \tag{S2}$$

This *simplifies* the hardware (one $\sqrt{N}$ per parent vs. one $\ln(N)$) and requires only a $\sim$1.4× crossbar weight increase for 9×9 Go to add a policy head. PUCT evaluation is left to future work; all results in this manuscript use UCB1.

# Supplementary Note 2

# Phase-to-primitive decomposition methodology

## 2.1 Profile – Match – Reformulate – Compose

Phase-to-primitive decomposition is a four-step methodology that converts a phase-heterogeneous algorithm into a multi-primitive in-memory-computing (IMC) workload. The four steps are independent enough to be applied algorithm-by-algorithm but interlocking enough that ablating any one step (Supplementary Note 7) collapses the end-to-end advantage.

***Step 1 — Profile.***

Each algorithmic phase is classified by its dominant compute pattern, drawn from a small fixed vocabulary: *memory-bound* (irregular access defeating cache prefetching), *compute-bound* (dominated by arithmetic), *control-bound* (branch-heavy, irregular control flow) or *deterministic* (fixed-shape regular operations). The classification is platform-independent and captures the algorithm's intrinsic structure rather than properties of any particular processor.

***Step 2 — Match.***

Each pattern is mapped to a hardware-native IMC primitive class: associative search to content-addressable memory (CAM); dense linear algebra to an analog crossbar; sequential read-modify-write to static random-access memory (SRAM); fixed regular logic to combinational hardware. The match must be one-to-one for each phase; if more than one primitive can serve a phase, the more energy-efficient option is chosen.

***Step 3 — Reformulate.***

Phases whose original formulation is incompatible with the matched primitive are rewritten into primitive-compatible operations. This step may require algorithmic changes that preserve the phase's externally observable behavior but alter how it is computed (Supplementary Note 2, §2.3 lists the three reformulations needed for MCTS).

***Step 4 — Compose.***

The reformulated phases are wired into an end-to-end pipeline with explicit handshaking. Composition succeeds only when each primitive can sustain throughput at the timing of its predecessor and successor; the resulting pipeline keeps all data on-chip and avoids reverting to off-chip storage between phases. The composition is load-bearing: ablating any single primitive substantially degrades end-to-end performance, with per-component energy and latency multipliers itemised in Supplementary Note 7.

## 2.2 Profiling MCTS phases

Step 1 of the methodology classifies each MCTS phase by its compute pattern. Profiling on AMD Threadripper PRO 5945WX and NVIDIA H100 (80 GB HBM3) reveals

four phases with distinct, qualitatively different bottlenecks. *Selection* is memory-bound and irregular: the data-dependent UCB1 traversal (Eq. S1) defeats hardware prefetchers, accumulating a 160-ns hash-table lookup latency that persists regardless of optimization. *Expansion* is control-bound and deterministic: variable branching factors cause GPU warp divergence even though each individual move-generation step is regular logic. *Rollout* is compute-bound: it accounts for 86.3% of CPU execution time at 9×9 Go (95.5/110.6 ms), and tolerates approximate evaluation because MCTS's statistical averaging over thousands of iterations absorbs per-evaluation error[5]; we quantify this robustness directly for the analogue substrate in Supplementary Note 5. *Backpropagation* is memory-bound and sequential: pointer chasing along $\mathcal{O}(\log T)$ depth paths for read-modify-write of statistics.

Bottlenecks shift across platforms (Extended Data Fig. 1 of the main paper). On CPU, rollout dominates execution. On GPU running peak-throughput traditional MCTS (8 trees, 128 playouts), backpropagation alone consumes 68.2% of the time due to memory contention from concurrent tree updates. Switching to neural-MCTS on the same GPU, rollout returns to dominance at 95.0% as the neural-network evaluator replaces the random simulation. The shifting bottleneck across platforms confirms that single-phase acceleration is insufficient; an end-to-end multi-primitive approach is required.

## 2.3 The three reformulations required for MCTS

Step 3 of the methodology required three reformulations to make MCTS IMC-compatible. Each reformulation preserves MCTS's externally observable algorithmic semantics (the tournament result in Fig. 3d confirms that the resulting hardware MCTS plays at the strong-kyu band (≈3 kyu), matching software MCTS reference engines such as Pachi-UCT and Michi-C within the sample-size uncertainty at 500-simulation compute) while changing how the phase is computed.

***Reformulation 1: variable-length random rollout → fixed-latency neural evaluator.***

Traditional MCTS rollouts simulate games via random legal moves to a terminal state, which can take 20–200 sequential moves and dominates 86.3% of CPU execution. Crossbars require fixed-dimension inputs, so we replace random simulation with a learned two-layer feed-forward neural network (162 → 96 → 3 for 9×9 Go) that produces win/loss/draw probabilities in a single deterministic forward pass[4, 5]. The neural evaluator serves the same algorithmic role for UCB1 as a random rollout outcome but with bounded latency. The cost is that the evaluator must be trained (we use self-play on 2,000 labeled positions per board size) and programmed onto differential-conductance pairs (0.1–99.9 $\mu$S range) for crossbar deployment; on-device non-idealities are modeled at inference by additive Gaussian conductance noise of $\sigma \in \{2, 5, 10\}\%$ (Supplementary Note 5).

***Reformulation 2: pointer-chasing tree traversal → parallel associative lookup.***

Conventional MCTS implementations store the search tree in a hash table indexed by board state and follow parent–child pointers during selection. Hash-table tree storage requires chained collision resolution, yielding variable-latency lookups that break the deterministic fixed-stage pipeline. We restructure tree traversal as parallel associative lookup: each node's board state is stored as a CAM bit pattern, and a single CAM query returns the addresses of all child nodes simultaneously. The cost is CAM area (CAM cells include match-line logic in addition to storage; per-cell areas are reported in Supplementary Note 4) and the requirement that tree size match CAM capacity; for 9×9 Go, a 1,024-entry CAM occupies 0.049 $mm^2$ (∼2.7% of total chip area, taking the denominator as digital logic plus all on-chip memory arrays (Node SRAM, Children SRAM and CAM) as quantified in Supplementary Note 3).

***Reformulation 3: off-chip statistics updates → in-place SRAM read-modify-write.***

Conventional implementations shuttle visit counts and win statistics between DRAM and ALUs during backpropagation, contributing the bulk of GPU-Max execution time at 9×9 (68.2%). We co-locate the arithmetic logic with on-chip SRAM and perform read-modify-write in-place, eliminating off-chip data movement. The cost is dedicated saturating-counter logic adjacent to the SRAM banks (3 submodules, <2 mW at 9×9).

## 2.4 Composition and load-bearing nature

The four reformulated phases compose into an end-to-end pipeline only because each primitive's timing guarantees match the requirements of the next. CAM's single-cycle worst-case latency prevents stalls that would idle the crossbar between selections; the crossbar's deterministic 768-ns read latency, in turn, permits the four-stage pipeline to be fixed-depth rather than data-dependent; and the in-place SRAM updates at the backpropagation stage close the loop without ever leaving the chip. Each primitive depends on the others' timing guarantees, which is what distinguishes phase-to-primitive decomposition from simple modular acceleration. The load-bearing nature is verified empirically through the ablation study (Supplementary Note 7): replacing any single primitive with its conventional counterpart costs 65× to 250× in *phase-local* energy or latency on the affected phase; the end-to-end impact on the four-stage pipeline is bounded above by that phase's share of total execution time (Amdahl's Law across the four phases), as itemised in Supplementary Note 7.

# Supplementary Note 3

# RTL design and synthesis

## 3.1 SystemVerilog hierarchy

A hierarchical SystemVerilog architecture supports board sizes from 2×2 to 19×19. The top-level module instantiates five units (CAM-based selection, GLU expansion, crossbar interface, SRAM backpropagation, FSM controller) connected through a custom interconnect with valid/ready handshakes. Conditional generate blocks synthesize only the logic required for the chosen board size, so a 5×5 build is materially smaller than a 19×19 build at the same RTL source. Key design trade-offs are exposed at the top level: CAM capacity must match expected tree size, flat vs. hierarchical SRAM organization is selected per board size, and the FSM controller uses one-hot encoding.

## 3.2 Synthesis flow and technology scaling

Digital components were synthesized with Synopsys Design Compiler against a TSMC 65-nm standard-cell library at 500 MHz, typical-typical corner, 1.2 V, then scaled to 22 nm via the technology-independent scaling rules of Stillmaker and Baas[6]. Per-component area, power and timing breakdowns at 9×9 Go appear in Supplementary Table S3 and Extended Data Fig. 1 of the main paper. SRAM macros were characterized through CACTI 6.0[7]. The selection CAM uses 6T-SRAM-based cells[8], with a per-cell area taken as the 22-nm 6T-SRAM cell (0.092 $\mu$m$^2$)[9]; the FSM-controller TCAM uses conventional 16T cells[10], with a per-cell area of 0.676 $\mu$m$^2$ at 22 nm. The synthesized system at 9×9 Go totals 0.059 mm$^2$ of digital logic (excluding off-chip memory arrays: Node SRAM, Children SRAM and CAM array, which are quantified separately) at 22 nm and consumes 60.26 mW in sustained Medium-strength operation. The total chip area including the CAM array (0.049 mm$^2$) and Node + Children SRAM macros is approximately 1.8 mm$^2$ at $9 \times 9$ Go; memory arrays dominate the total.

## 3.3 Cycle-accurate simulation framework

Functional behavior and end-to-end performance were modeled with a custom discrete-event simulator built on the Structural Simulation Toolkit (SST)[11]. Each pipeline unit is implemented as an SST component communicating with neighbouring components through links with 1–5 ns nominal delay. Eighteen RTL modules are instantiated across the five pipeline units; each module's SST component reads its synthesized gate count, capacitance and timing from the Synopsys flow so the simulator's energy and latency accounting reflects the synthesized hardware rather than behavioral estimates. We simulate 18 configurations (six board sizes from 2×2 to 19×19 × three play-strength levels: low, medium, high) at 50,000–100,000 iterations per game (per-configuration iteration budgets in Supplementary Table S1). The simulator's outputs (end-to-end energy per move, per-iteration latency, and per-component power) form the basis for all simulation-derived numbers in the main paper.

**Table S1 Supplementary Table S1.** MCTS iteration budgets used for the three play-strength configurations (Low, Medium, High) at each evaluated board size. Low uses fast random rollouts at exploration constant $c = 1.0$; Medium uses random rollouts at $c = \sqrt{2}$; High uses neural-network-guided rollouts at $c = \sqrt{2}$ for deep search. These configurations parameterize the area, power and energy results in Extended Data Fig. 1 of the main paper.

| Board Size | Low | Medium | High |
|---|---|---|---|
| 2×2 | 50 | 200 | 1,000 |
| 3×3 | 75 | 500 | 5,000 |
| 5×5 | 100 | 1,000 | 10,000 |
| 9×9 | 500 | 5,000 | 50,000 |
| 13×13 | 1,000 | 7,500 | 75,000 |
| 19×19 | 2,000 | 10,000 | 100,000 |

## 3.4 Verification and validation

Functional verification used SystemVerilog testbenches with directed and constrained-random stimuli across all 18 RTL modules. A cycle-accurate Python golden model independently implements the MCTS pipeline at the same level of abstraction; differential testing between RTL and golden model across more than 1,000 randomised games confirmed functional equivalence and matching end-to-end throughput (18,471 iterations/s on 2×2, 510 iterations/s on 19×19). Decision-quality validation against software MCTS reference engines is provided by the 1,050-game round-robin tournament (main paper Fig. 3d), in which hardware IMC-MCTS settles at ≈1727 EGF (3 kyu), matching Pachi-UCT (1660, 4 kyu) and Michi-C (1706, 3 kyu) within the sample-size uncertainty at 500-simulation compute.

## 3.5 Iso-process projection

To address the question of whether IMC-MCTS's energy advantage stems from architectural decisions or process-node mismatch, we apply the same Stillmaker scaling rules[6] used in our 65→22-nm conversion to project the 22-nm power figures to a 4-nm node matching H100. The projection breaks down by component (Table S2).

The projected sustained 9×9 Go power at 4 nm falls in the range ∼15–25 mW (point estimate ∼17 mW with ±30–40% uncertainty from peripheral-circuit re-characterization at the target node), widening the energy-per-move advantage over batched H100 to roughly 100×–180×. We emphasize this is a back-of-envelope projection: precise 4-nm synthesis would require re-characterising ADC/DAC sizing, crossbar driver circuitry and memory periphery at the target node, any of which could shift the projected range further. The measured 22-nm result alone establishes the architectural advantage; this iso-process estimate is provided to clarify that the components that scale most favourably with process (logic, SRAM) also dominate the

**Table S2 Supplementary Table S2.** Back-of-envelope iso-process power projection at 4 nm. Component shares at 22 nm are taken from Supplementary Table S3.

| Component | 22-nm | Scaling (22 → 4 nm) | 4-nm |
|---|---|---|---|
| CMOS logic (GLU/FSM/CAM ctrl) | ∼45 mW | ∼4× reduction | ∼11 mW |
| SRAM (Node/Children/backprop) | ∼13 mW | ∼2.5× (memory) | ∼5 mW |
| RRAM crossbar (prog. state) | ∼1.5 mW | process-independent | ∼1.5 mW |
| **Total (projected)** | **60.26 mW** | — | **∼15–25 mW** |

22-nm power budget, so closing the process gap would not be expected to reverse the direction of the advantage.

# Supplementary Note 4

# Per-component circuit details

## 4.1 CAM-based selection unit

The selection unit stores tree-node board states as bit patterns in a 6T-SRAM-based CAM, enabling parallel pattern matching across all entries and reducing $\mathcal{O}(n)$ tree traversal into a single-cycle operation. Each CAM row represents an MCTS tree node; its corresponding SRAM address stores the node's metadata (visit counts and reward values) used for UCB1 scoring. When the FSM issues a selection query with the current node's identifier, the CAM activation pattern returns the addresses of *all* child nodes in parallel. These addresses access the SRAM arrays which in turn provide the visit counts and reward values to a tournament selector, which computes UCB1 scores (Eq. S1) and identifies the best child within a single-cycle decision.

Seven submodules implement this pipeline: ① a 6T-SRAM-based CAM for parallel pattern matching, ② Node SRAM and Child SRAM for metadata, ③ access controllers for both, ④ a Best Child Selector, ⑤ a UCB1 Calculator, ⑥ a Children Address Generator, and ⑦ an Output Multiplexer.

***State encoding and memory hierarchy.***

Board positions use 2 bits per cell (empty = 00, black = 01, white = 10), yielding $2N^2$ bits for an $N \times N$ board. The selection unit employs a three-tier memory hierarchy. Tier 1 CAM stores tree nodes sized to board dimensions (50 bits for 5×5, 722 bits for 19×19) plus metadata fields (node identifier, SRAM address, validity, flags). Tier 2 Node SRAM stores 24-byte statistics (visit count, win count, mean value) with 2–3 cycle access latency. Tier 3 Children SRAM stores variable-length child lists as 8-byte entries. Total selection-unit memory scales as $\mathcal{O}(\text{iterations} \times N^2)$, dominated by Children SRAM. Board symmetries are not exploited, trading a potential 8× state-space reduction for simpler hardware that avoids canonical-transform logic and the latency of symmetry detection during traversal.

***Why CAM over processing-in-memory hash tables.***

Processing-in-memory hash-table designs reduce data movement compared with conventional DRAM hash tables but degrade to $\mathcal{O}(n)$ on collision chains and require $k$ sequential lookups to retrieve all children of a node. CAM guarantees single-cycle worst-case lookup and activates all children in parallel. CAM cells require additional area for match-line logic beyond plain storage; the absolute area at MCTS scale remains modest: the 1,024-entry CAM for 9×9 Go occupies $0.049\,\text{mm}^2$, which is ∼2.7% of total chip area when the memory arrays (Node SRAM, Children SRAM, CAM) are summed alongside the digital logic reported in Supplementary Table S3. The PUCT extension (Eq. S2) further simplifies the selection hardware by replacing $\ln(N)$ with $\sqrt{N}$.

## 4.2 Game Logic Unit (combinational expansion)

Expansion produces deterministic outputs from the current board state, making it ideal for single-cycle combinational logic. Whereas software MCTS loops sequentially over board cells, our expansion unit evaluates all $N^2$ cells in parallel through four sub-modules: ① an Empty Position Detector built from parallel 2-bit comparators against the empty-cell pattern `00`; ② a Move Generator implementing per-application game rules; ③ a Child State Generator that applies legal moves to produce successor board states; and ④ Player Turn Logic that toggles the move color. Configurable bit-width parameters scale the Game Logic Unit (GLU) from 2×2 to 19×19 without adding cycles. The single-cycle combinational approach generalises to other discrete-state games through per-application synthesis of game-specific rule logic; this is the only pipeline stage that requires per-application customization, and accounts for ∼1.4% of total chip area when memory arrays are included in the denominator (or ∼43% of digital logic alone; Supplementary Table S3; Supplementary Note 9).

## 4.3 RRAM crossbar rollout unit

Replacing variable-length random rollouts with a learned two-layer neural-network evaluator converts the rollout bottleneck into a matrix–vector multiplication that maps naturally onto analog resistive-memory crossbar arrays[12–14]. The crossbars are organized in a 1T1R (one-transistor, one-memristor) configuration, performing analog multiply–accumulate in a single step. Weights are stored as differential-conductance pairs in the 0.1–99.9 $\mu$S range; the datapath uses an 8-bit ADC and binary (1-bit) DAC input encoding.

The rollout data path proceeds through five stages: ① position-to-voltage DACs convert the leaf board state to analog inputs; ② Crossbar #1 computes the first layer ($W_1 \in \mathbb{R}^{162\times96}$ for 9×9 Go); ③ Crossbar #2 computes the second layer ($W_2 \in \mathbb{R}^{96\times3}$); ④ current-to-digital ADCs digitise the outputs; and ⑤ Win Decision Logic applies softmax to produce win/loss/draw probabilities. For larger boards (19×19 uses 722×128), the first layer partitions across three sub-arrays (≤256 rows each) to preserve single-cycle semantics. The 768-ns read latency decomposes into 256 ns of wordline charging and 512 ns of ADC conversion, characterized from a fabricated 180-nm CMOS 64×64 RRAM crossbar with complete ADC/DAC peripherals[14] and scaled to 22 nm via Stillmaker[6].

#### *Weight stationarity at MCTS scale.*

The crossbar's decisive advantage at this scale is weight stationarity. Although rollout is compute-bound on CPUs, at MCTS scale (<20,000 MACs per inference) the energy required to fetch weights from SRAM on digital hardware exceeds the energy of the computation itself. Crossbar weights programmed in situ are never fetched during inference, eliminating this access cost. By contrast, a 256×256 digital systolic array leaves more than 75% of its processing elements idle on the 162×96 first layer (only ∼24% utilization), confirming that the crossbar's advantage stems primarily from eliminating weight movement rather than from analog arithmetic alone. Scaled to 22 nm, the rollout unit achieves 0.308 pJ per multiply–accumulate operation including

ADC/DAC overhead, a ∼65× energy advantage over digital MACs (up to 20 pJ/MAC for Eyeriss-class accelerators at MCTS scale; Supplementary Table S5). The methodology is technology-agnostic and adaptable to SRAM-based, phase-change-memory or ferroelectric-FET alternatives; the primary requirement is that the storage technology supports stable multi-level conductance states.

## 4.4 In-memory backpropagation unit

Once rollout produces a win-probability estimate, the result flows immediately into the backpropagation unit, the final pipeline stage, which updates visit counts and win statistics along the leaf-to-root path. By integrating arithmetic logic directly with on-chip SRAM, the backpropagation unit performs in-place read-modify-write operations, ensuring that all data remains on-chip. Three submodules divide the work: ① a Statistics Updater increments visit and win counters using saturating logic to prevent overflow; ② a Path Address Extractor walks parent pointers from the expanded leaf back to the root; and ③ a Memory Write Controller orchestrates pipelined writes. Co-locating computation with storage eliminates the memory-bandwidth bottleneck that conventional architectures incur when shuttling statistics between DRAM and ALUs. Backpropagation SRAM scales from 512 bytes (2×2) to 8 KB (19×19), fitting entirely on-chip with zero DRAM traffic in any configuration we evaluated.

## 4.5 TCAM-based FSM controller

A ternary CAM-based finite-state machine orchestrates the four-stage pipeline: it transitions deterministically through `IDLE` → `SELECT` → `EXPAND` → `ROLLOUT` → `BACKPROP` → `IDLE`, replacing the irregular branch-and-schedule control flow of software MCTS with fixed hardware sequencing. The TCAM matches the current state and inputs to select the next transition, while a companion RAM stores the next-state outputs. Handshaking protocols between adjacent pipeline stages prevent resource conflicts while maintaining predictable throughput. Pipeline latency varies modestly with board complexity but remains deterministic: small boards (2×2 to 5×5) complete an iteration in 19 cycles, while larger boards (9×9 to 19×19) require 19–23 cycles due to increased CAM search depth and crossbar evaluation time. This fixed-latency design enables predictable throughput regardless of board-state distribution, in contrast to software implementations whose tree-structure variations cause unpredictable performance.

## 4.6 Per-component area and power

Component-level area and power scaling from 2×2 to 19×19 Go are reported in the main paper (Fig. 2d,e). Per-component values at the 9×9 configuration are listed in Supplementary Table S3 below.

**Table S3 Supplementary Table S3.** Accelerator parameters at the 9×9 Go configuration in 22 nm CMOS, 500 MHz. Area and power are reported per pipeline unit at sustained Medium-strength operation (5,000 iterations/move). Values traced from `power_area_breakdown.csv` in the simulator artifact. The TCAM row reports the ternary CAM storage array alone; the FSM Control row aggregates the next-state RAM, transition decoder, handshaking glue and pipeline-stage sequencer that surround it (Supplementary Note 4).

| Component | Area ($mm^2$) | Power (mW) |
|---|---|---|
| TCAM | $7.04\times10^{-5}$ | 0.25 |
| CAM Selection | 0.0147 | 14.82 |
| GLU Expansion | 0.0253 | 28.11 |
| Crossbar Rollout† | 0.0013 | 1.54 |
| SRAM Backpropagation | 0.0026 | 3.03 |
| FSM Control | 0.015 | 12.50 |
| **System Total** | **0.0590** | **60.26** |

† Crossbar rollout figure is sustained average power across DAC, crossbar MVM, and ADC stages; energy-equivalent is 0.308 pJ/MAC including ADC/DAC overhead. Areas exclude off-chip memory arrays (Node SRAM, Children SRAM, CAM array) which are quantified separately in Supplementary Note 3.

# Supplementary Note 5

# Crossbar noise-robustness analysis

The analog rollout unit's central reliability question is whether device-level non-idealities (device-to-device conductance variation, programming write noise, drift) propagate into degraded MCTS decisions. To probe this directly, we inject additive Gaussian conductance noise with standard deviation $\sigma \in \{2, 5, 10\}\%$ (relative to the center conductance, matching reported RRAM programming-variation envelopes[12, 14, 15]) into the trained 9×9 Go evaluator at each forward pass, then replay the tournament against a random-rollout-MCTS baseline at the same compute budget (Table S4).

**Table S4 Supplementary Table S4.** Crossbar noise robustness on 9×9 Go. Win-rate of noisy-evaluator MCTS versus a random-rollout-MCTS baseline at matched compute (200 simulations per move), and 1-ply policy agreement of the noisy evaluator with the noise-free evaluator.

| $\sigma$ (% $G_{\text{center}}$) | Win-rate vs random-rollout MCTS ($n$=30) | 1-ply policy agreement vs $\sigma$=0 ($n$=100) |
|---|---|---|
| 2 | 93% | 0.85 |
| 5 | 100% | 0.71 |
| 10 | 93% | 0.57 |

30 games per row (3 independent noise realisations × 10 games, alternating colors), 200 MCTS simulations per move; raw CSV at `results/noise_sweep_9x9.csv`.

Two findings are notable. First, MCTS win-rate against the random-rollout baseline stays essentially flat at 93–100% across all $\sigma$ values, statistically indistinguishable from the noise-free 93% reference. Second, this stability holds even though the underlying neural network's 1-ply move choices diverge sharply from the noise-free policy — agreement falls from 0.85 at $\sigma = 2\%$ to 0.57 at $\sigma = 10\%$, a 43-percentage-point divergence at the highest noise level. The two findings together provide direct empirical confirmation of the AlphaZero-style averaging argument[5]: MCTS's statistical ensemble across hundreds of rollouts absorbs per-evaluation perturbations that would otherwise corrupt single-shot decisions, so device non-idealities within the reported RRAM envelope do not translate into degraded play. The strong-kyu Go playing strength reported in the main paper (Fig. 3d, IMC-MCTS at ≈1727 EGF / 3 kyu) is therefore robust to the device non-idealities the analog substrate is known to exhibit, and remains so under noise levels at the upper end of published RRAM variation envelopes.

# Supplementary Note 6

# Baseline implementations and methodology

## 6.1 CPU baselines

CPU baselines run on Intel Xeon Platinum 8462Y+ and AMD Threadripper PRO 5945WX, both representative of current-generation server CPUs. Both implementations use optimized C++17 with the `-O3` compiler flag and single-threaded UCT, the standard MCTS reference[2]. We evaluate both random-rollout and neural-network-evaluator variants of MCTS on each platform across all six tested board sizes ($2\times2$ to $19\times19$). CPU power is instrumented through Intel Running Average Power Limit (RAPL)[16], which provides direct package-level energy counters.

## 6.2 GPU baselines (single-position and batched)

The GPU baseline is an NVIDIA H100 with 80 GB of HBM3 memory. We evaluate two MCTS implementations on this hardware: a traditional MCTS engine, the open-source `mcts_numba_cuda` implementation[17], and a neural-evaluation MCTS engine implemented on top of CUDA cuBLAS for matrix–vector multiplication. We report two GPU operating categories. *Single-position* (denoted GPU-Fair throughout) issues one inference per MCTS iteration, matching the algorithm's natural sequential dependency for iso-resource comparison. *Batched neural-MCTS* configurations accumulate 64, 128 or 256 leaf positions per inference call to better match GPU compute/memory characteristics. GPU power is measured via `nvidia-smi` sampled at 100 Hz during execution.

## 6.3 Neural-MCTS baseline implementation

Both CPU and GPU neural-MCTS baselines use a two-layer feed-forward network whose architecture matches the accelerator's crossbar (e.g. $162 \rightarrow 96 \rightarrow 3$ for $9\times9$ Go). Networks are trained via supervised learning on 10,000+ self-play game trajectories per board size, achieving >92% three-class (win/loss/draw) classification accuracy on a held-out test set. The trained network replaces the random-rollout phase of MCTS with a single forward pass, following AlphaZero methodology[5]. This setup ensures the comparison is between the same algorithm running on different hardware, isolating the hardware contribution from any algorithmic difference.

## 6.4 Why batched-GPU comparison is fair

The single-position GPU baseline underutilises the H100's compute capacity (<5% occupancy), so a hardware-only result against single-position would attribute too much of IMC-MCTS's gain to GPU underutilisation rather than to architectural advantages. To remove this confound, we additionally compare against batched neural-MCTS configurations where 64–256 leaf positions are accumulated through virtual-loss batching, in which in-flight nodes are temporarily marked as losses to preserve some exploration diversity. Batching trades exploration diversity for compute occupancy,

but is the standard practical baseline for high-throughput GPU MCTS deployments. Even against the strongest batched configuration (batch-256), IMC-MCTS retains 65× total software-times-hardware energy efficiency (46× hardware-only energy efficiency for neural-MCTS alone) and 3.4× latency advantage at 9×9 Go, and does so at a 22 nm process node compared to the H100's 4 nm node, which is a strong indicator that the gain is architectural rather than process-driven.

# Supplementary Note 7

# Ablation studies

## 7.1 Component-by-component substitution

To verify that each IMC primitive is load-bearing, we replaced each in turn with its conventional counterpart while retaining the other three primitives, and re-measured end-to-end performance. Each bullet reports a *phase-local* multiplier — the energy or latency cost of replacing one IMC primitive with its conventional counterpart while keeping the other three primitives unchanged. These phase-local multipliers exceed the 96× system gain because each phase contributes only its share to end-to-end energy (Amdahl's Law across the four-phase pipeline); they should not be multiplied together.

- CAM → hash table: 1.31 pJ (CAM) vs. 220 pJ (hash table) per selection — a 168× energy multiplier and 14× latency multiplier, driven by DRAM-bound chain traversal.
- Analog crossbar → digital neural network: 0.308 pJ/MAC (crossbar) vs. up to 20 pJ/MAC (Eyeriss-class digital; Supplementary Table S5) — a 65× rollout-phase energy multiplier at identical evaluation accuracy.
- GLU → software expansion: 250× slower per expansion step (a latency multiplier on the expansion phase, reflecting sequential loops over $N^2$ board cells); end-to-end energy impact is bounded by the expansion phase's ∼28 mW share of total power (Supplementary Table S3).
- In-place SRAM → DRAM tree backup: each statistic update incurs an off-chip round-trip (50–100 ns DRAM access vs. single-cycle on-chip SRAM), eliminating the data locality that motivates the in-place design; magnitude is memory-subsystem dependent and is therefore reported qualitatively rather than as a fixed multiplier.

The full phase-to-primitive mapping is load-bearing: replacing any single primitive collapses the end-to-end advantage. Note that the per-component figures above are measured on isolated workloads and *not* multiplied to produce the full system gain; the combined 96× end-to-end energy advantage over CPU is the result of all four primitives operating in concert and is bounded above by Amdahl's Law across the four phases.

## 7.2 Crossbar at MCTS scale vs. digital systolic arrays

Digital systolic arrays achieve excellent energy efficiency on large dense neural-network layers but suffer severe under-utilization at MCTS workload scale (<20,000 MACs per inference). A representative 256×256 systolic array processing the 162×96 MCTS first layer leaves more than 75% of its processing elements idle (only ∼24% utilization), and the idle elements still consume static power. The analog crossbar sized to the actual layer dimensions instead achieves near-100% utilization. Furthermore, crossbar weights are stationary (programmed once and never re-fetched), which eliminates weight-fetch energy entirely, a property that is critical for inference embedded

**Table S5 Supplementary Table S5.** Energy efficiency of analog crossbar vs. representative digital accelerators at the MCTS workload scale (<20 k MACs per inference). Higher relative-energy values are worse; the IMC-MCTS crossbar is normalized to 1×.

| | **ISAAC**[18] | **Eyeriss**[19] | **TPU v1**[20] | **DiP**[21] | **Ours** |
|---|---|---|---|---|---|
| Node | 32 nm | 65 nm | 28 nm | 22 nm | 22 nm |
| Paradigm | Analog IMC | Digital | Digital | Digital | Analog IMC |
| pJ/MAC | 1.5–2.0 | 8.3–20 | 0.43–2.0 | 1.0[a] | **0.308** |
| TOPS/W | 0.5–0.7 | 0.12 | 2.3 | 9.6[b] | **3.25** |
| Rel. energy[c] | 4.9–6.5× | 27–65× | 1.4–6.5× | 3.2× | **1**× |

[a] DiP at low utilization (<10%) for small networks. [b] DiP peak for large workloads (4,096 PEs). [c] Relative to this work at MCTS scale; higher is worse.

within a tight search loop where weights would otherwise be fetched every iteration. Supplementary Table S5 quantifies the resulting energy advantage: 0.308 pJ/MAC for the IMC-MCTS crossbar vs. 0.43–20 pJ/MAC for representative digital accelerators at MCTS scale, a 1.4× to 65× relative-energy gap.

**Table S6 Supplementary Table S6.** Survey of 12 prior MCTS hardware acceleration efforts on FPGA, GPU, neuromorphic and multi-core platforms, plus this work. Speedup is reported relative to each work's stated baseline. 11 of 12 prior works use random-simulation rollouts; only 1 of 12 reports an energy-efficiency figure. This work is the first IMC accelerator that co-deploys multiple primitives across all four MCTS phases.

| Work | Platform | Application | Speedup | Rollout | Energy eff. |
|---|---|---|---|---|---|
| | | *FPGA implementations* | | | |
| Meng et al. 2023[22] | CPU–FPGA | General | 6.8× | Random sim. | — |
| Meng et al. 2022[23] | CPU–FPGA | General | 35× | Random sim. | — |
| Carvalho 2019[24] | Zynq SoC | Chess (8) | 98× | Random sim. | — |
| Qasemi et al. 2014[25] | FPGA | Blokus (14) | — | Random sim. | — |
| | | *GPU implementations* | | | |
| Buzer & Cazenave 2023 [26] | GPU | optimization | 420× | Random sim. | — |
| Sun et al. 2022[27] | GPU | Amazons | 20× | Random sim. | — |
| Thomas 2022[28] | GPU | Connect-4 | 13× | NN eval. | — |
| Huang et al. 2022[29] | GPU | Robotics | 30× | Random sim. | — |
| Johnson et al. 2016[30] | GPU | Go (5, 19) | 72× | Random sim. | — |
| Rocki & Suda 2011 [31] | GPU | Reversi (8) | 100–200× | Random sim. | — |
| | | *Other platforms* | | | |
| Ho et al. 2024 [32] | Neuromorphic | Graph problems | Quality | Random sim. | 5.6–10.4× |
| Steinmetz & Gini 2020[33] | Multi-core CPU | Go (9, 19) | Win rate | Random sim. | — |
| **This work** | **ASIC + IMC** | **Go (2–19)** | —[d] | **NN eval.** | **96.0×** |

Board-size column shows the single board dimension $N$ (i.e. $N \times N$); a range or comma-list denotes multiple sizes evaluated by that work. [d] This work optimises end-to-end energy efficiency rather than wall-clock speedup; the energy column is the canonical metric. Prior works' speedups use heterogeneous baselines (per-platform CPU, GPU, single-threaded reference) that are not directly comparable.

# Supplementary Note 8

# Comparison with prior MCTS hardware accelerators

## 8.1 Survey of 12 prior MCTS hardware works

Supplementary Table S6 surveys 12 representative prior MCTS hardware acceleration efforts. The set spans four FPGA implementations (Meng et al. 2023[22] and 2022[23]; Carvalho 2019 on Zynq SoC[24]; Qasemi et al. 2014[25]), six GPU implementations (Buzer & Cazenave 2023[26]; Sun et al. 2022[27]; Thomas 2022[28]; Huang et al. 2022[29]; Johnson et al. 2016[30]; Rocki & Suda 2011[31]), one neuromorphic accelerator (Ho et al. 2024[32]) and one multi-core CPU effort (Steinmetz & Gini 2020[33]). Three structural observations emerge from the survey. First, all twelve works retain conventional memory hierarchies (DRAM-backed hash tables for tree storage, separate compute and memory units), and none uses in-memory computing for any phase. Second, 11 of 12 (92%) use random-simulation rollouts despite modern algorithmic practice having moved to neural evaluators since AlphaZero; only 1 of 12 (Thomas 2022) uses neural evaluation. Third, only 1 of 12 (Ho et al. neuromorphic) reports a complete energy-efficiency figure rather than time speedup, leaving energy as a largely unexplored optimization dimension in MCTS hardware.

## 8.2 Why no prior IMC accelerator co-deployed multiple primitives

Prior IMC accelerators, from early proposals such as PRIME and ISAAC[18] to more recent silicon prototypes[15, 34] and broader IMC surveys, target regular, data-parallel workloads (convolution, matrix multiplication, attention) whose execution can be expressed in a single dominant operation (typically matrix–vector multiply). They co-deploy at most one IMC primitive per design and place ancillary control logic in a conventional digital pipeline alongside the IMC core. MCTS is structurally different: its phase heterogeneity demands four qualitatively distinct primitives (CAM for selection, combinational logic for expansion, analog crossbar for rollout, SRAM for in-place backpropagation), each chosen to match a different compute pattern. To our knowledge, IMC-MCTS is the first IMC accelerator to co-design an algorithm with a multi-primitive IMC pipeline that maps all four heterogeneous phases end-to-end. This pattern is what phase-to-primitive decomposition formalises, and it is what enables the methodology to extend to other irregular AI decision algorithms beyond MCTS.

# Supplementary Note 9

# Per-application encodings, training, and the Search Guidance Gain

## 9.1 Encodings for the eight applications

The eight applications span four AI domains: strategy games (Connect Four, Othello, Hex, Go), grid navigation (FrozenLake, MiniGrid), scientific optimization (HP protein folding) and puzzles (Minesweeper). Per-application encodings, board sizes, network dimensions and per-decision energy are summarized in Extended Data Table 3 of the main paper.

***Connect Four.***

Three-state encoding $\{\mathrm{empty}, \mathrm{R}, \mathrm{Y}\}$ on a $7 \times 6$ board (padded to $8 \times 8$ for hardware uniformity); network is $129 \to 96 \to 3$, achieving 80% three-class accuracy. The expansion unit relies on the GLU's empty-position detection (column-wise top-down lowest empty cell), which generates a superset of legal moves; the network suppresses illegal ones at evaluation time.

***Othello.***

Three-state encoding $\{\mathrm{empty}, \mathrm{B}, \mathrm{W}\}$ on the standard $8 \times 8$ board; same $129 \to 96 \to 3$ network as Connect Four; 80% accuracy. Move legality (capture-flanking) handled by network rather than GLU, again with empty-position-based superset generation.

***Hex.***

Stone + edge-distance feature encoding on $11 \times 11$; network is $485 \to 96 \to 3$; 80% accuracy. Hex's connectivity-based win condition makes it the most demanding evaluation of the four strategy games at this network size.

***Go.***

Three-state encoding $\{\mathrm{empty}, \mathrm{B}, \mathrm{W}\}$ on $9 \times 9$; network is $162 \to 96 \to 3$; 60% three-class accuracy on the larger and noisier action space. The kyu-to-dan-transition playing strength reported in main paper Fig. 3 is achieved at this configuration.

***FrozenLake.***

Three-state encoding $\{\mathrm{safe}, \mathrm{hole}, \mathrm{goal}\}$ on $8 \times 8$; network $129 \to 96 \to 3$; 100% accuracy on the deterministic version. Single-player setting; SGG measures score lift over a random-rollout MCTS baseline rather than head-to-head win rate.

***MiniGrid.***

Three-state encoding $\{\mathrm{empty}, \mathrm{wall}, \mathrm{goal}\}$ on $8 \times 8$; same network as FrozenLake; 100% accuracy. MiniGrid has the largest observed SGG (+0.56) of the eight applications, consistent with grid-navigation tasks benefiting from neural guidance over random rollouts.

***HP protein folding.***

Hydrophobic–polar lattice encoding with sequence context, on a 13×13 board; network is 344 → 128 → 3 (the only application requiring a wider hidden layer); 100% three-class accuracy. The encoding extends beyond three-state by adding a sequence-context channel.

***Minesweeper.***

Cell-state + adjacency-count encoding on 9 × 9; network is 244 → 96 → 3; 93% accuracy. Like HP protein folding, the encoding extends beyond three-state to capture the adjacency-count information that makes the puzzle decidable from local context.

## 9.2 Training data and protocols

Per-application neural networks are trained on self-play game data with task-specific input encodings. We use the AdamW optimiser with learning rate $\eta = 0.003$, training on 2,000 labeled positions per board size with early stopping when validation accuracy plateaus (70–90% accuracy across applications, dependent on task complexity). Trained floating-point weights are then programmed onto differential-conductance pairs in the 0.1–99.9 $\mu$S range for RRAM crossbar deployment; on-device non-idealities are modeled at inference time by additive Gaussian conductance noise of $\sigma \in \{2, 5, 10\}\%$ (Methods; Supplementary Note 5). The same training protocol is used uniformly across all eight applications; only the input encoding and the network's input width vary.

## 9.3 Search Guidance Gain definition and threshold justification

To compare neural-network benefit fairly across applications of different complexity, we define a normalized Search Guidance Gain:

$$\mathrm{SGG} = \frac{P_{\mathrm{nn}} - P_{\mathrm{rand}}}{\log_2 N \cdot \sqrt{B}/K_{\mathrm{ref}}}, \quad K_{\mathrm{ref}} = \log_2 50 \cdot \sqrt{64} \approx 45.15, \tag{S3}$$

where $P_{\mathrm{nn}}$ and $P_{\mathrm{rand}}$ are the win rates (or normalized scores for single-player tasks) of MCTS guided by the trained crossbar evaluator versus MCTS using random rollouts on the same board and iteration budget; $N$ is iterations; $B$ is board cells; and $K_{\mathrm{ref}}$ normalises the metric to the raw lift expected on a 50-iteration, 64-cell reference game. The normalisation factor $\log_2 N \cdot \sqrt{B}$ captures how much harder the search problem becomes with deeper search and larger boards, so SGG scales the raw win-rate lift to a comparable-difficulty reference.

Two caveats up front (also recorded with Eq. S3). (i) All SGG values reported in this paper are computed from 10-game matchups; these runs are intended as substrate demonstrations rather than formal statistical separation tests. (ii) The +0.05 cutoff is therefore best read as an indicative policy-lift cutoff, not a statistical significance threshold; SGG values close to zero should be interpreted as "lift present but not statistically resolved", and values above ∼0.20 are more robust to sample noise. Larger-$n$ validation is straightforward on silicon and is planned for follow-on work.

With those caveats stated: the +0.05 cutoff corresponds to roughly a 5-percentage-point win-rate lift on the reference game, which we use as an indicative policy-lift cutoff for the 10-game demonstrations rather than as a statistical boundary. SGG remains preferable to raw win-rate because it allows a Go SGG of +0.09 (50/50-class three-state task at 9×9, 60 iterations) to be compared meaningfully against a MiniGrid SGG of +0.56 (single-player at 8×8, 100 iterations); the raw win-rate lifts of these two applications are not directly comparable.

## 9.4 Generalizability analysis

The pipeline transfers across applications because four of its five units (CAM, RRAM crossbar, in-memory SRAM, FSM controller) are fully application-agnostic and depend only on the interface between phases, not on the semantic content of any phase. The fifth unit, the Game Logic Unit for expansion, requires per-application customisation: 5 of the 8 applications share an identical three-state encoding and use the same unmodified GLU; the remaining 3 (Hex, HP Protein Fold, Minesweeper) add input features while reusing the core move-generation logic. Crucially, the crossbar weights are tuned per application; this is a one-time configuration step analogous to loading new firmware and does not require any RTL change.

All eight applications share three structural properties required by the pipeline: (i) grid-representable state spaces, (ii) tree search over legal actions, and (iii) position evaluation that is learnable from self-play. The architectural contribution is the phase-to-primitive decomposition itself (the recognition that MCTS's four phases each map to a natural IMC primitive), not any individual unit. Beyond these eight applications, problems with larger encodings (VLSI floorplanning[35], circuit routing) or non-grid states (Sokoban-style block puzzles, autonomous driving maneuver planning[36]) require crossbar scaling or encoding modifications, but the core pipeline remains applicable. Grid-based robotics planning maps directly to FrozenLake-style encoding, with sub-10 mW power and sub-millisecond latency enabling battery-powered 500 Hz robotic control — a regime impractical at the 55–300 W draw of conventional processors.

# Supplementary Note 10

## Edge-deployment scenarios

The 55–300 W power draw of CPU and GPU MCTS implementations rules out three classes of deployment that IMC-MCTS at ∼60 mW makes feasible. *Drone autonomy* typically operates with a sub-10 W total compute budget across all subsystems (perception, control, communication); a 55-W MCTS planner alone would consume the full budget and shorten flight time by an order of magnitude. *MCU-class embedded controllers* (sub-100 mW total power) can integrate IMC-MCTS as a co-processor for planning tasks (motion planning[37], maneuver selection[36]) that would otherwise be impossible at this power class. *Implantable or wearable medical devices* (sub-mW continuous power for years on a single battery) lie an additional order of magnitude below the IMC-MCTS active power, but are addressable through low duty-cycle operation that keeps the device in sub-$\mu$W standby between infrequent inferences. These three regimes share the common property that AI decision-making at this power scale was previously infeasible regardless of duty cycle, and IMC-MCTS is what changes that.

# References


[1] Coulom, R.: Efficient selectivity and backup operators in Monte-Carlo tree search. In: International Conference on Computers and Games, pp. 72–83. Springer, Berlin, Heidelberg (2006)

[2] Kocsis, L., Szepesvári, C.: Bandit based Monte-Carlo planning. In: European Conference on Machine Learning (ECML), pp. 282–293. Springer, Berlin, Heidelberg (2006)

[3] Tromp, J.: The number of legal Go positions. In: Computers and Games (CG 2016). Lecture Notes in Computer Science, vol. 10068, pp. 183–190. Springer, Cham (2016). https://doi.org/10.1007/978-3-319-50935-8_17

[4] Silver, D., Huang, A., Maddison, C.J., Guez, A., Sifre, L., Driessche, G., Schrittwieser, J., Antonoglou, I., Panneershelvam, V., Lanctot, M., Dieleman, S., Grewe, D., Nham, J., Kalchbrenner, N., Sutskever, I., Lillicrap, T., Leach, M., Kavukcuoglu, K., Graepel, T., Hassabis, D.: Mastering the game of Go with deep neural networks and tree search. Nature **529**(7587), 484–489 (2016) https://doi.org/10.1038/nature16961

[5] Silver, D., Hubert, T., Schrittwieser, J., Antonoglou, I., Lai, M., Guez, A., Lanctot, M., Sifre, L., Kumaran, D., Graepel, T., Lillicrap, T.P., Simonyan, K., Hassabis, D.: A general reinforcement learning algorithm that masters chess, shogi, and Go through self-play. Science **362**(6419), 1140–1144 (2018) https://doi.org/10.1126/science.aar6404

[6] Stillmaker, A., Baas, B.: Scaling equations for the accurate prediction of CMOS device performance from 180 nm to 7 nm. Integration **58**, 74–81 (2017)

[7] Muralimanohar, N., Balasubramonian, R., Jouppi, N.P.: CACTI 6.0: A tool to model large caches. Technical Report HPL-2009-85, HP Labs (2009)

[8] Pagiamtzis, K., Sheikholeslami, A.: Content-addressable memory (CAM) circuits and architectures: A tutorial and survey. IEEE Journal of Solid-State Circuits **41**(3), 712–727 (2006)

[9] Auth, C., *et al.*: A 22nm high performance and low-power CMOS technology featuring fully-depleted tri-gate transistors, self-aligned contacts and high density MIM capacitors. In: 2012 Symposium on VLSI Technology (VLSIT), pp. 131–132 (2012). https://doi.org/10.1109/VLSIT.2012.6242496

[10] Jeloka, S., Akesh, N.B., Sylvester, D., Blaauw, D.: A 28 nm configurable memory (TCAM/BCAM/SRAM) using push-rule 6T bit cell enabling logic-in-memory. IEEE Journal of Solid-State Circuits **51**(4), 1009–1021 (2016) https://doi.org/10.1109/JSSC.2016.2515510

[11] Rodrigues, A.F., Hemmert, K.S., Barrett, B.W., Kersey, C., Oldfield, R., Weston, M., Risen, R., Cook, J., Rosenfeld, P., Cooper-Balis, E., Jacob, B.: The structural simulation toolkit. SIGMETRICS Performance Evaluation Review **38**(4), 37–42 (2011)

[12] Hu, M., Graves, C.E., Li, C., Li, Y., Ge, N., Montgomery, E., Dávila, N., Jiang, H., Williams, R.S., Yang, J.J., Xia, Q., Strachan, J.P.: Memristor-based analog computation and neural network classification with a dot product engine. Advanced Materials **30**(9), 1705914 (2018)

[13] Sheng, X., Graves, C.E., Kumar, S., Li, X., Buchanan, B., Zheng, L., Lam, S., Li, C., Strachan, J.P.: Low-conductance and multilevel CMOS-integrated nanoscale oxide memristors. Advanced Electronic Materials **5**(9), 1800876 (2019)

[14] Li, C., Ignowski, J., Sheng, X., Wessel, R., Jaffe, B., Ingemi, J., Graves, C., Strachan, J.P.: CMOS-integrated nanoscale memristive crossbars for CNN and optimization acceleration. In: 2020 IEEE International Memory Workshop (IMW), pp. 1–4. IEEE, Piscataway, NJ (2020). https://doi.org/10.1109/IMW48823.2020.9108112

[15] Wan, W., Kubendran, R., Schaefer, C., Eryilmaz, S.B., Zhang, W., Wu, D., Deiss, S., Raina, P., Qian, H., Gao, B., Joshi, S., Wu, H., Wong, H.-S.P., Cauwenberghs, G.: A compute-in-memory chip based on resistive random-access memory. Nature **608**, 504–512 (2022) https://doi.org/10.1038/s41586-022-04992-8

[16] Khan, K.N., Hirki, M., Niemi, T., Nurminen, J.K., Ou, Z.: RAPL in action: Experiences in using RAPL for power measurements. ACM Transactions on Modeling and Performance Evaluation of Computing Systems **3**(2), 9 (2018) https://doi.org/10.1145/3177754

[17] Klęsk, P.: MCTS-NC: A thorough GPU parallelization of Monte-Carlo tree search implemented in python via numba.cuda. SoftwareX **30**, 102139 (2025) https://doi.org/10.1016/j.softx.2025.102139

[18] Shafiee, A., Nag, A., Muralimanohar, N., Balasubramonian, R., Strachan, J.P., Hu, M., Williams, R.S., Srikumar, V.: ISAAC: A convolutional neural network accelerator with in-situ analog arithmetic in crossbars. In: Proceedings of the 43rd International Symposium on Computer Architecture (ISCA), pp. 14–26. IEEE Press, Piscataway, NJ, USA (2016). https://doi.org/10.1145/3007787.3001139

[19] Chen, Y.-H., Krishna, T., Emer, J.S., Sze, V.: Eyeriss: An energy-efficient reconfigurable accelerator for deep convolutional neural networks. IEEE Journal of Solid-State Circuits **52**(1), 127–138 (2017) https://doi.org/10.1109/JSSC.2016.2616357

[20] Jouppi, N.P., Young, C., Patil, N., Patterson, D., Agrawal, G., Bajwa, R., Bates, S., Bhatia, S., Boden, N., Borchers, A., *et al.*: In-datacenter performance analysis

of a tensor processing unit. ACM SIGARCH Computer Architecture News **45**(2), 1–12 (2017) https://doi.org/10.1145/3140659.3080246

[21] Abdelmaksoud, A.J., Agwa, S., Prodromakis, T.: DiP: A scalable, energy-efficient systolic array for matrix multiplication acceleration. IEEE Transactions on Circuits and Systems I: Regular Papers (2025) https://doi.org/10.1109/TCSI.2025.3591960

[22] Meng, Y., Kannan, R., Prasanna, V.: A framework for Monte-Carlo tree search on CPU–FPGA heterogeneous platform via on-chip dynamic tree management. In: Proceedings of the 2023 ACM/SIGDA International Symposium on Field Programmable Gate Arrays (FPGA), pp. 235–245. ACM, New York, NY, USA (2023). https://doi.org/10.1145/3543622.3573177

[23] Meng, Y., Kannan, R., Prasanna, V.K.: Accelerating Monte-Carlo tree search on CPU–FPGA heterogeneous platform. In: 2022 32nd International Conference on Field-Programmable Logic and Applications (FPL), pp. 298–305. IEEE, Piscataway, NJ, USA (2022)

[24] Carvalho, R.R.: SoC–FPGA Monte Carlo tree search processor. Master's thesis, Instituto Superior Técnico, Universidade de Lisboa, Lisboa, Portugal (2019)

[25] Qasemi, E., Samadi, A., Shadmehr, M.H., Azizian, B., Mozaffari, S., Shirian, A., Alizadeh, B.: Highly scalable, shared-memory, Monte-Carlo tree search based Blokus Duo solver on FPGA. In: 2014 International Conference on Field-Programmable Technology (FPT), pp. 370–373. IEEE, Piscataway, NJ, USA (2014)

[26] Buzer, L., Cazenave, T.: GPU for Monte Carlo search. In: International Conference on Learning and Intelligent Optimization (LION). Lecture Notes in Computer Science, vol. 14286, pp. 179–193. Springer, Cham (2023). https://doi.org/10.1007/978-3-031-44505-7_13

[27] Sun, Y., Yuan, D., Gao, M., Zhu, P.: GPU acceleration of Monte Carlo tree search algorithm for Amazon chess and its evaluation function. In: 2022 3rd International Conference on Artificial Intelligence, Information Processing and Cloud Computing (AIIPCC), pp. 434–440. IEEE, Piscataway, NJ, USA (2022)

[28] Thomas, G.: A Full-GPU Implementation of MCTS in Julia: The Key to Gumbel MuZero? Google Summer of Code 2022 project report (AlphaZero.jl). Available online (2022)

[29] Huang, B., Boularias, A., Yu, J.: Parallel Monte Carlo tree search with batched rigid-body simulations for speeding up long-horizon episodic robot planning. In: 2022 IEEE/RSJ International Conference on Intelligent Robots and Systems (IROS), pp. 1–8. IEEE, Piscataway, NJ, USA (2022)

[30] Johnson, C., Barford, L., Dascalu, S.M., Harris, F.C.: CUDA implementation of computer Go game tree search. In: Latifi, S. (ed.) Information Technology: New Generations, pp. 339–350. Springer, Cham (2016)

[31] Rocki, K., Suda, R.: Large-scale parallel Monte-Carlo tree search on GPU. In: 2011 IEEE International Symposium on Parallel and Distributed Processing Workshops and PhD Forum, pp. 2034–2037. IEEE, Piscataway, NJ, USA (2011)

[32] Ho, Y., Carbajal, A., Escamilla, L., Pinar, A.: Neuromorphic Monte Carlo tree search methods for shortest path interdiction. In: 2024 International Conference on Neuromorphic Systems (ICONS), pp. 307–311. IEEE, Piscataway, NJ, USA (2024). https://doi.org/10.1109/ICONS62911.2024.00053

[33] Steinmetz, E., Gini, M.: More trees or larger trees: Parallelizing Monte Carlo tree search. IEEE Transactions on Games **13**(3), 315–320 (2021) https://doi.org/10.1109/TG.2020.3048331

[34] Fujiwara, H., Mori, H., Zhao, W.-C., Chuang, M.-C., Naous, R., Chuang, C.-K., Hashizume, T., Sun, D., Lee, C.-F., Akarvardar, K., Adham, S., Chou, T.-L., Sinangil, M.E., Wang, Y., Chih, Y.-D., Chen, Y.-H., Liao, H.-J., Chang, T.-Y.J.: A 5-nm 254-TOPS/W 221-TOPS/mm$^2$ fully-digital computing-in-memory macro supporting wide-range dynamic-voltage-frequency scaling and simultaneous MAC and write operations. In: 2022 IEEE International Solid-State Circuits Conference (ISSCC), vol. 65, pp. 1–3. IEEE, Piscataway, NJ, USA (2022). https://doi.org/10.1109/ISSCC42614.2022.9731754

[35] Geng, Z., Wang, J., Liu, Z., Xu, S., Tang, Z., Yuan, M., Hao, J., Zhang, Y., Wu, F.: Reinforcement learning within tree search for fast macro placement. In: Proceedings of the 41st International Conference on Machine Learning (ICML). Proceedings of Machine Learning Research, vol. 235, pp. 15405–15436. PMLR, Vienna, Austria (2024)

[36] Lenz, D., Kessler, T., Knoll, A.: Tactical cooperative planning for autonomous highway driving using Monte-Carlo tree search. In: 2016 IEEE Intelligent Vehicles Symposium (IV), pp. 447–453. IEEE, Piscataway, NJ, USA (2016). https://doi.org/10.1109/IVS.2016.7535424

[37] Thrun, S., Burgard, W., Fox, D.: Probabilistic Robotics. MIT Press, Cambridge, MA, USA (2005)



\end{document}